\documentclass{JHEP3}

\input epsf
\usepackage{amsmath}
\usepackage{amssymb}
\usepackage{graphicx}

\newcommand{\ads}[1]{\mbox{AdS}_{{#1}}}

\newcommand{\adscft}{{$\ads{}$/CFT}}

\newcommand{\half}{\frac12}

\newcommand{\beq}{\begin{eqnarray}}
\newcommand{\eeq}{\end{eqnarray}}

\def\mn{{\mu\nu}}

\def\be{\begin{equation}}
\newcommand{\bel}[1]{\be\label{#1}}
\def\ee{\end{equation}}
\newcommand{\eref}[1]{(\ref{#1})}
\newcommand{\Eref}[1]{Eq.~(\ref{#1})}
\newcommand{\rem}[1]{}

\def\half{{1\over 2}}
\def\NN{{\cal N}}

\def\elll{\ell}

\def\ntwo{$\NN=2$}

\def\nfour{$\NN=4$}

\def\bit{\begin{itemize}}
\def\eit{\end{itemize}}

\def \be {\begin{equation}}
\def \ee {\end{equation}}
\def \bea {\begin{eqnarray}}
\def \eea {\end{eqnarray}}
\def \half{\frac{1}{2}}
\newcommand{\ket}[1]{{\left | {#1}\right \rangle}}
\newcommand{\bra}[1]{{\left \langle {#1}\right |}}
\newcommand{\q}{\bar{q}}

\newcommand{\matt}[1]{{}}
\newcommand{\cut}[1]{{}}
\newcommand{\move}[1]{{}}

\newcommand{\ylm}{{\cal Y}^{\ell}}

\title{On the Couplings
of Vector Mesons in AdS/QCD}

\author{Sungho Hong$^{ab}$, Sukjin Yoon$^{a}$,
and Matthew J. Strassler$^{a}$\\
$^{a}$Department of Physics and Astronomy\\
P.O Box 351560, University of Washington\\
Seattle, WA 98195\\
\\
$^{b}$Department of Physics and Astronomy\\
University of Pennsylvania\\
Philadelphia, PA 19104-6396}


\abstract{We address, in the \adscft\ context, the issue of the
universality of the couplings of the $\rho$ meson to other hadrons.
Exploring some models, we find that generically the $\rho$-dominance
prediction $f_\rho g_{\rho H H}=m_\rho^2$ does not hold, and that
$g_{\rho H H}$ is not independent of the hadron $H$.  However, we
prove that, in any model within the  AdS/QCD context, 
there are two limiting regimes where the
$g_{\rho H H}$, along with the couplings of all excited vector mesons
as well, become $H$-independent: (1) when $H$ is created by an
operator of large dimension, and (2) when $H$ is a highly-excited
hadron.   We also find a sector of
a particular model where universality for the $\rho$ coupling is
exact.  Still, in none of these cases need it be true that $f_\rho
g_\rho=m_\rho^2$, although we find empirically
that the relation does hold approximately (up to a factor of
order two) within the models we have studied.}

\preprint{
hep-th/0409118\\
UW/PT 04-16\\
UPR-1090-T}

\begin{document}

\section{Introduction}
The observed couplings of the octet of
vector mesons ($\rho(770)$, $\omega(782)$ ,
$\phi(1020)$, etc.) show an interesting universality, one which is not
an obvious consequence of any known QCD mechanism. The $\rho$ decay
to $\pi\pi$
gives $g_{\rho\pi\pi}^2/(4\pi)=2.9$, and isospin-related
decays of the $\phi$ give
$g_{\phi K^+ K^-}^2/(4\pi)=3.2$ and $g_{\phi K_L
  K_S}^2/(4\pi)=3.5$~\cite{klingl}; this should be compared
with the unrelated process of pion-nucleon scattering, which yields
$g_{\rho \pi\pi}
g_{\rho NN}/(4 \pi) = 2.8$~\cite{sakurai66}.

In 1960, Sakurai proposed a now-famous conjecture~\cite{sakurai60},
that the $\rho$ meson has a universal coupling to every
isospin-carrying hadron.  In particular, the ``vector meson
dominance'' conjecture~\cite{gell-mann,sakurai,williams} sets this
universal coupling to the $\rho$ mass-squared divided by the $\rho$
decay constant: $g_\rho=m_\rho^2/f_\rho$.  The suggestion is that
the form factor of any isospin-carrying hadron $H$ is given by
the $\rho$ pole:
$$
F(q^2) \approx {f_\rho g_{\rho H H}\over q^2+m_\rho^2}
$$ 
where $f_\rho$ is the $\rho$ decay constant, $m_\rho$ is its mass,
and $g_{\rho H H}$ is a coupling characterizing the interaction of a
$\rho$ with the hadron $H$.  The $H$-independent normalization
condition $F(0)=1$ then fixes $g_\rho$.  Sakurai attempted to
implement this idea by formulating the $\rho$ meson as a gauge boson.
This approach was influential and inspired much subsequent work.

It seems to us that a modern
viewpoint, particularly employing the techniques of
\adscft~\cite{Maldacena}, might shed some interesting light on
this old issue.  
One of our motivations in exploring this question is a recent
proposal by Son and Stephanov based on dimensional
deconstruction~\cite{SonSteph}, which in turn was inspired by
the hidden-local symmetry mechanism of Bando {\it et al.} \cite{bando}  
An interesting
aspect of this model is
that $\rho$-dominance for some hadrons is
a natural consequence of the properties of wavefunctions in the
deconstructed extra dimension.  There has been a similar observation
in another model based on \adscft~\cite{hys}.
In this theory, the form factors associated with certain conserved
global symmetry
currents 
are expressed in terms of only a {\em
finite} sum of poles; these poles correspond to the vector meson
states created by the current acting on the vacuum.  
As in~\cite{SonSteph}, the
special properties of the extra-dimensional mode functions play an
essential role.  Since the theory of \cite{SonSteph} is an {\it ad hoc} model,
constructed by hand, it is useful to see that the same mathematics
arises in an \adscft\ context, where the whole structure of the
computation, including the extra dimension, 
arises naturally as the dual picture of a
strongly-coupled field theory.

In this paper, we will examine
the universality of the $\rho$'s couplings, and those of other
excited vector mesons created by the same current.\footnote{We will
generically call the lowest-mass state created by a conserved current
acting on the vacuum the ``$\rho$'', at the risk of some confusion.}  
We find the universality and $\rho$-dominance conjectures do 
not hold across entire
models, both in that couplings
are nonuniversal and do not satisfy $f_\rho g_{\rho H H} = m_\rho^2$\ .  
However,   we do find that universal couplings for {\it all} of the
vector mesons emerge, model-independently,
in two interesting limits. 
These two examples of approximate universality hold for
diagonal couplings of the vector mesons 
both to hadrons whose interpolating
operator has large dimension and to hadrons which have high radial
excitation. As we will show, the existence of universal
couplings in these limits is a 
consequence of general properties of the \adscft\
calculation.
We also find an example of exact universality, whose origin is
interesting but clearly model-dependent, 
for the couplings
of the $\rho$ to the hadrons within  a large sector of
a particular model.
In this example the 
universal coupling is of the same order as, but does not equal,
$m_\rho^2/f_\rho$.

The paper is organized as follows. In section~\ref{sec:decomp}, we
discuss vector meson dominance in the limit of 
large $N$ and large 't Hooft
coupling; a proof of vector meson dominance in this limit,
due to Son, is given in appendix~\ref{sec:proof}.  
We will supplement the discussion by examples in two different
models, which are reviewed in appendix~\ref{sec:review-models}.  
One is the ``hard-wall'' model, which is used to capture
generic features of confining gauge theories.  The other model is the
D3/D7 system, which has ``quarks'' in the fundamental representation,
and associated strongly-coupled ``quarkonium'' bound states. In
section~\ref{sec:universality}, we briefly discuss
$\rho$-dominance and how it motivates the study of coupling
universality. Then, we lay out the various types
of universality that we have explored, illustrating them with
examples from the hard-wall and  D3/D7
models.  Section~\ref{sec:concl-disc} contains some concluding remarks. A
review of the basic methodology needed from the AdS/CFT dictionary
can be
found in appendix~\ref{sec:methodology}.

\section{Decomposition in \adscft}\label{sec:decomp}

In the literature on hadronic physics, it is often assumed that the
form factor for a hadron associated with a conserved spin-one current
can be written as a sum over vector-meson poles. While this
is not in general justified, it is believed to be true at large $N$.
\matt{Discuss with Son one more time}
The main goal of this section is to argue this is
indeed always true in \adscft\ contexts, when both the number of
colors $N$ and the 't Hooft coupling $\lambda=g^2N$ are large.  
Here $g$ is
the Yang-Mills coupling.

In particular, we claim
(and sketch a proof, due to Son \cite{son}, in
appendix~\ref{sec:proof}) that
in confining gauge theories with a supergravity dual,\footnote{There 
can be many form factors
depending on whether the current is conserved, and on the spins of the
hadrons. In this paper, we only deal with conserved currents,
whose matrix elements between scalar hadrons 
have only one form factor.
For vector hadrons, there are three form factors: electric $F_e$,
magnetic $F_m$ and quadrupole $F_Q$. As observed by Son and Stephanov
\cite{SSprivate} and discussed in \cite{hys}, the
large 't Hooft coupling limit implies $F_e=F_m$ and $F_Q=0$.
Therefore, our discussion focusing on only one form factor is justified.}
\begin{equation}\label{Ffg}
\quad F_{ab} ( q^2 ) = \sum_n \frac{f_n g_{n ab}}{q^2 + m_n^2},
\end{equation}
as illustrated in Fig.~\ref{fig:decomp}.
Here $f_n$ denotes the hadron decay constant of the $n$-th vector
hadron state, $g_{n ab}$ its coupling to an incoming and outgoing
hadron, and $m_n$ its mass. 
\FIGURE[ht]{
\epsfbox{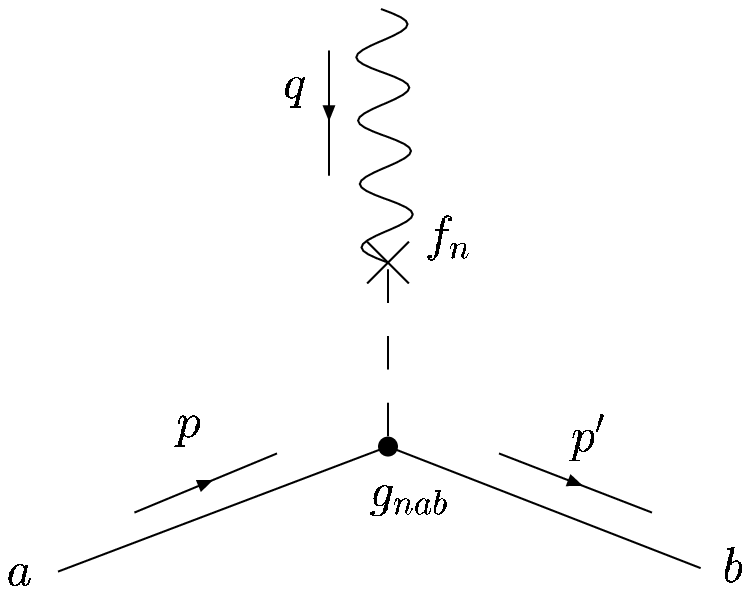}
\caption{Decomposition of the form factor into a sum over hadron
states, as in Eq.~(\ref{Ffg}).}
\label{fig:decomp}}

Before we begin, we need to review how the form factor is computed on
the gravity side of \adscft. Leaving the details to
appendix~\ref{sec:methodology}, we cover only what is needed in this
section. According to the \adscft\ duality, a local conserved spin-one
current in the gauge theory is dual to a {\em non-normalizable} mode
of a gauge field 
in the asymptotically-$\ads{5}$ space.  Meanwhile,
the spin-one 
hadron state created by the current
operator corresponds to a {\em normalizable} mode of the same
gauge field.  Now, the form
factor is computed by the overlap integral, \Eref{eq:formfactor}, 
of a non-normalizable mode
and two normalizable modes, corresponding to the vector current, an
incoming hadron, and an outgoing hadron.  The three hadron coupling,
in which the current is replaced by a spin-one hadron created
by that current, is obtained
by the {\em same integral} except for the 
replacement of the non-normalizable mode 
of the five-dimensional gauge field with a normalizable one, as in
\Eref{eq:THO}.\footnote{For
spin-two currents, the same statements hold with all spin-one currents
and hadrons replaced with spin-two, and with the five-dimensional
gauge field replaced by the five-dimensional graviton.}
Therefore, the decomposition~(\ref{Ffg}), if true, must be
derived simply from a relationship between the normalizable and
non-normalizable modes. 

The required relationship is the following.
Let's consider a 
spin-$J$
($J\leq2$) field living in the asymptotically $\ads{5}$ space
(which we will assume is embedded in a
$d$-dimensional asymptotically-
$\ads{5}\times W$ space, with
$W$ a compact manifold of dimension $d-5$.)
The mode of the field
with momentum $q^\mu$
may be written $C_{\mu_1 \cdots \mu_J}(q)
 = \epsilon_{\mu_1
\cdots \mu_J} e^{iq\cdot x} \chi(q^2,z)$,
where $z$ is the five
dimensional radial coordinate defined in 
appendix~\ref{sec:methodology}, and $\mu$
runs from 0 to 3. 
The normalizable mode is
$\phi_n(z)\propto\chi(-m_n^2,z)$ 
at $q^2=-m_n^2$, and the
non-normalizable mode $\psi(q^2,z)\equiv\chi(q^2,z)$ for arbitrary
$q^2$ can be written as
\begin{equation}\label{eq:drel}
\psi(q^2,z)=\sum_n \frac{f_n \phi_n(z)}{q^2+m_n^2}
\end{equation}
\begin{equation}\label{eq:fn}
f_n=\lim_{z\to 0}
\frac{V(z)}{g_d}\left(\frac{R}{z}\right)\ \partial_z
\phi_n(z) \ .
\end{equation}
Here $g_d$ is a $d$-dimensional coupling constant; its
precise form depends on the current and the theory under study.
Meanwhile $R=\lambda^{1/4}\alpha'^{1/2}$ is the $\ads{5}$ curvature radius,
and 
$V(z)$ is the volume of $W$ at $z$.
Substitution of \Eref{eq:drel} into
(\ref{eq:formfactor}) and using (\ref{eq:THO}) yields \Eref{Ffg}.  A
proof of~(\ref{eq:drel}) and \eref{eq:fn} for spin-one currents is
given in appendix~\ref{sec:proof}.  Additional details about our
notation are given in appendix~\ref{sec:methodology}.

It will be useful below to recall the scaling properties
of the $f_n$ in models with supergravity duals.
The decay constant $f_n$ of a spin-one hadron is defined
by
\[ \bra{0} \mathcal{J}^{\mu} (x=0) \ket{n, p, \epsilon } = f_n
 \epsilon_\nu,
\]
where $\ket{0}$ is the vacuum of the theory, and $\ket{n, p,
  \epsilon}$ is the spin-one hadron state with mass $m_n$, momentum
  $p$ and polarization $\epsilon_{\mu}$ created by the conserved
  current operator $\mathcal{J}^{\mu}$. 
The $f_n$ and $m_n$ are constrained by the fact that the
two-point correlation function of a conserved current can be written
\begin{equation}\label{eq:twocor}
\left \langle \mathcal{J}_{\mu} ( q ) \mathcal{J}_{\nu} ( - q )
\right\rangle 
\sim q^2 \ln q^2 
\left(\eta_{\mu\nu}-\frac{q_\mu q_\nu}{q^2}\right) = 
\left ( q^2 \eta_{\mu \nu} - q_{\mu}
   q_{\nu} \right) \sum_n \frac{|f_n|^2}{m_n^2(q^2 + m_n^2)} .
\end{equation}
If $m_n\sim n^p$ at large $n$, then the $\log q^2$ behavior requires
$f_n\sim n^{2p-1/2}$.  The supergravity limit has $p=1$, so $f_n\sim
n^{3/2}$ for a conserved current.  Similarly, the energy momentum
tensor has $f_n\sim n^{5/2}$ since its two-point function goes as
$q^4\ln q^2$.  We know of no similarly useful constraints on the
three-hadron couplings $g_{nab}$, unless all three hadrons are highly
excited, a case we will not discuss.

\subsubsection*{Examples}

We now illustrate 
the above formalism through a few examples. Our computation will be mostly
focused on, first, showing the decomposition~(\ref{Ffg}) explicitly in
two exactly solvable models, and second, verifying the formula for the
hadron decay constant $f_n$~(\ref{eq:fn}). The computed $f_n$'s will
be also useful in comparing the examples in future sections with
previous work.

The first theory we consider is the hard-wall model, which effectively
models the behavior of a large class of confining theories in
the large $\lambda$ limit. On the gravity side, the theory is simply given by
an $\ads{5}\times S^5$ background cut off by a wall at finite radius, where
boundary condition for mode functions are imposed.
We leave the detailed discussion of this theory to the original
literature \cite{hardscat,polchinskisusskind,DIS}, but the list of mode
functions that we will use can be found in
appendix~\ref{sec:review-models}.

The other model that we use is the flavor-non-singlet sector of a
system of $N$ D3-branes and $N_f$ D7 branes.  This model has a
distinctive feature; the theory has QCD-like mesons, bound states
built from matter in the fundamental representation of $SU(N)$. The
meson spectrum has been largely worked out. We again refer to
appendix~\ref{sec:review-models} for a brief introduction to the
theory and for the mode functions; the reader may wish to consult the
original literature~\cite{KK,KKW,Myers,hys} for a more detailed
description of the theory.

\paragraph{\underline{Hard-wall Model}}

In the hard-wall model, we recall the
non-normalizable~(\ref{eq:Ahadnorm}) and normalizable
mode~(\ref{Ahadwave}) for a gauge field in this theory,
\[
\psi (q,z) \approx \frac{1}{g_{10}}
q z K_1 (q z)\quad\text{(non-normalizable mode)},
\]
\[
\phi_n (z) = \frac{\sqrt{2}\Lambda z J_1(\zeta_{0,n}\Lambda z)}{
\pi^\frac32
R^3
  J_1(\zeta_{0,n})}\quad\text{(normalizable mode)}.
\]
where $\zeta_{\nu,n}$ is the $n^{th}$ zero of $J_\nu$, and the
approximation in the first equation is that $q\gg \Lambda$.
As explained in
  appendices~\ref{sec:methodology} and \ref{sec:review-models}, 
the canonical normalization of the
  non-normalizable mode requires division by $g_{10} = \kappa/R$, the
  effective coupling constant for a spin-one mode in the hard-wall
  model. %
These two functions are related by a mathematical identity,
\begin{equation}\label{eq:besselKJ}
q^\nu K_\nu ( q x ) = \int^{\infty}_0 d m \,
   \frac{m^{\nu+1} J_\nu ( m x )}{q^2 + m^2}.
\end{equation}
For $\nu=1$, this formula is of the same form as
Eq.~(\ref{eq:drel}), except that the sum over states has
been replaced with an integral over a continuous spectrum.  
The reason for this is that 
in constructing the
non-normalizable mode as in~(\ref{eq:Ahadnorm}) we ignored the
boundary condition on the wall, which
leads us to a continuous spectrum.  This spectrum
approximates the
true discrete result in the limit of high-mass states ($n\gg 1$)
or equivalently in the limit of small confinement scale $\Lambda$.

Comparing Eq.~(\ref{eq:drel}) with (\ref{eq:besselKJ}), we see $f_n$
  is given simply by $m_n^2$ divided by the normalization coefficient
  of the normalizable mode. 
\be\label{eq:so4fn}%
f_n = \frac{d m_n}{d n} m_n^2 \left( {\sqrt{2} \Lambda \over 
\pi^\frac32 
R^3 J_1 (
    m_n / \Lambda )}\right)^{-1} \left( \frac{R}{\kappa} \right)
{\approx} \frac{
\pi^2}{2\sqrt{2}} n^2 J_1 ( \zeta_{0;n} )
\Lambda^{2} N \approx \frac\pi2 n^{3/2} \Lambda^2
N
\ee
The powers of $N$ and $\Lambda$ in this result
are fixed on
general grounds by $N$-counting and dimensional analysis.  The
$n^{3/2}$ scaling is required by ultraviolet conformal invariance, as
explained earlier.

Meanwhile Eq.~(\ref{eq:fn}) applied directly
to the normalizable mode $\phi_n$ gives
\[
f_n =
\lim_{z\to 0}
\frac{R}{\kappa}\cdot\frac{\pi^3R^6}{z}\cdot
\partial_z  \frac{\sqrt{2}\Lambda z J_1(\zeta_{0;n}\Lambda
    z)}{\pi^\frac32 R^3 J_1(\zeta_{0;n})}
= \frac{N\Lambda^{2} \zeta_{0;n}}{\sqrt{2} \pi
J_{1}(\zeta_{0;n})}
\approx \frac\pi2 
n^{3/2} \Lambda^2 N
.
\]
As we noted earlier, the discrepancy between these equations arises
from the fact that \Eref{eq:besselKJ} is exact only in the strictly
conformal limit $\Lambda \to 0$; the reader is invited to check that
the discrepancy is removed when the exact form of the non-normalizable
mode, \Eref{eq:Ahadnorm}, is used. As required, the two expressions match
in the large $n$ limit.

For future comparison, we also compute the ratio between $m_n^2$ and
$f_n$,
\begin{equation}\label{eq:gnhardwall}
\frac{m_n^2}{f_n} = \half \zeta_{0;n}
J_1(\zeta_{0;n})\frac{(2\pi)\sqrt{2}
}{
N}.
\end{equation}
In particular, for the $\rho$ ($n=0$),
\begin{equation}\label{eq:grhohardwall}
\frac{m_\rho^2}{f_\rho}=0.624 \frac{(2\pi)\sqrt{2}
}{
N} \ .
\end{equation}

The extension to the energy-momentum tensor
is straightforward. It corresponds to the $\nu=2$ case in
Eq.~(\ref{eq:besselKJ}) and the decomposition is explicit. Once
again, we also read off the decay constant of a spin two hadron,
\[
f_{n}  = \frac{d m_{n}}{d n} \left( \frac{m_n^3}{2} \right) 
\left( \frac{\sqrt{2} \kappa \Lambda}{\pi^\frac32 R^4
    J_{2}(m_{n}/\Lambda)}\right)^{-1} \approx \frac{\pi^{3}}{4
  \sqrt{2}} n^3 J_2 ( n \pi ) \Lambda^{3}N \sim \frac{\pi^2}4 n^{5 / 2}
\Lambda^{3} N.
\]
Eq.~(\ref{eq:fn}) gives
\[
f_{n} = \lim_{z\to 0}\frac1\kappa \cdot \frac{\pi^3 R^6}z \partial_z
\frac{\sqrt{2} \Lambda J_2(\zeta_{1;n}z)}{\pi^\frac32 R^2 J_2(\zeta_{1;n})} =
\frac{\zeta_{1;n}^2 \Lambda^{3} N}{ 2 \sqrt{2} \pi
  J_2(\zeta_{1;n})} \approx \frac{\pi^2}4 n^{5 / 2} \Lambda^{3} N.
\]
Again we observe coincidence in the conformal limit.

\paragraph{\underline{D3/D7 Model}}

Let us now turn to the D3/D7 system. It has been shown that the
decomposition~(\ref{Ffg}) is explicit in this case~\cite{hys}. Hence,
we will check only the formulas for $f_n$~(\ref{eq:fn}) in the
following ways: first, we compute $f_n$ for the cases where the form
factors are known explicitly, and second, we compare the result with
the one from Eq~(\ref{eq:fn}). We will also read off $f_n$ from the
form factors using different external hadrons; as we will see, the
$f_n$'s are independent of the external hadrons, as they should be.

Using the metric and the mode functions given in
appendix~\ref{sec:review-models}, the coupling constant is obtained by
an overlap integral for the type I modes and the vector mode, expressed
as
\begin{equation}\label{eq:typeIoverlap}
g^\ell_{n, n_1 , n_2} = g_8 \frac{L^2}{2} (2\pi^2) \int_0^1 \frac{dv}{ v^{2}}\,
\phi^{II}_{0,n} (v) \phi^{I}_{\ell,n_1} (v) \phi^{I}_{\ell,n_2} (v),
\end{equation}
where $g_8$ is the Yang-Mills coupling of the eight dimensional D7
worldvolume theory and $L$ is the distance between the D7 and D3
branes, which sets the quark mass $m_Q=L/\alpha'$. The typical meson
mass scale is set by $m_h=L/R^2=m_Q/\sqrt{\lambda}$. We compute a
special case of the vector $(0,n)$ -- scalar $(1,n_2)$ -- scalar
$(1,0)$ overlap integral,
\begin{eqnarray}
 g^{\ell=1}_{n,0,n_2} &=& (-1)^{n+n_2+1} \frac{(2\pi)}{\sqrt{N}}
 \sqrt{\frac{3}{2 (n+1)(n+2)(2n+3)(2n_2+3)}} \nonumber \\  
&\ &\times \left[n_2(n_2+2)\delta_{n,n_2-1}-(2n_2+3)\delta_{n,n_2}-(n_2+1)(n_2+3)
 \delta_{n,n_2+1}\right]\ .  
\label{eq:scalarTHO}
\end{eqnarray}
Comparing this with the form factor computed in~\cite{hys}, we obtain
the decay constant $f_n$ of the vector meson:\footnote{Note that we have
used a slightly different overall
normalization in this paper compared to \cite{hys}; $f_n$ and $g_n$
both differ by a factor of $2\pi^2$, the volume of a unit 3-sphere.  The
change cancels in form factors where only $f_ng_n$ appears.}
\begin{equation} f_n = (-1)^{n} \frac{m_{h}^2
\sqrt{N}}{(2\pi)}\sqrt{8(n+1)(n+2)(2n+3)} \sim m_{h}^2
n^{3/2}\sqrt{N} ,\label{eq:flavorhdc}
\end{equation}
where we used $(L^2 / R^4)(R^2/g_8) = m_{h}^2\sqrt{2N}/(2\pi)^2$.
The $n^{3/2}$ scaling for large $n$ is required by conformal
invariance, while the powers of $N$ and $m_h$ are fixed on general
grounds by $N$-counting and dimensional analysis.

We may now cross-check this result.
The type II normalizable mode with $\ell=0$ is
\begin{eqnarray*}
\phi^{II}_{n} (\varrho) & = &
\frac{C_{0n}^{II}/R^2}{(1 + \varrho^2 )^{n+1}} F(-n, -1-n ; 2 ; - \varrho^2 )\\
& = &  \frac{C_{0n}^{II}}{R^2} \frac{(-1)^n}{\varrho^2} + O\left(
\frac{1}{\varrho^3} \right) \ .
\end{eqnarray*}
Using this and Eq.~(\ref{eq:fn}), we obtain
\begin{eqnarray}
f_n & = & (2\pi^2/g_8) \rho^3 \partial_{\rho}{\phi^{II}_{n}}(\rho) |_{\rho
  \rightarrow \infty} \nonumber\\
& = & (-1)^{n} (2\pi)^2 (L^2 /g_8 R^2)C_{0n}^{II}\nonumber\\
& = & (-1)^{n} \frac{m_{h}^2 \sqrt{N}}{(2\pi)}
\sqrt{8(n+1)(n+2)(2n+3)},
\end{eqnarray}
which is exactly Eq.~(\ref{eq:flavorhdc}). 
Note that (in analogy to 
Eq.~(\ref{eq:gnhardwall}))
\begin{equation}\label{eq:gnd3d7}
\frac{m_n^2}{ f_n}=(-1)^n\sqrt{\frac{2(n+1)(n+2)}{2n+3}} 
\frac{(2\pi)}{\sqrt{N}}
\end{equation}
and for the $\rho$ ($n=0$)
\begin{equation}\label{eq:grhod3d7}
 \frac{m_\rho^2}{f_\rho}=\frac2{\sqrt{3}}
\frac{(2\pi)}{\sqrt{N}} \ .
\end{equation}

For comparison, we compute the three vector hadron coupling also.  It
is given by almost the same integral as (\ref{eq:typeIoverlap}), except
that $\phi^I_{\ell,n_i}\to \phi^{II}_{\ell,n_i}$; also the metric factor
changes accordingly:
\begin{equation}
  \label{eq:FFF}
  g^\ell_{n,n_1,n_2} = g_8 \frac{R^4}{2}(2\pi^2) 
\int_0^1 dv\,\left( \frac{1-v}{v}
\right) \phi^{II}_{0,n}(v) \phi^{II}_{\ell,n_1}(v) \phi^{II}_{\ell,n_2}(v).
\end{equation}
Now the vector $(0,n)$ -- vector $(0,n_2)$ -- vector $(0,0)$ overlap
integral is
\begin{multline}
 g^{\ell=0}_{n,0,n_2} = (-1)^{n+n_2+1} \frac{(2\pi)}{\sqrt{N}}
\sqrt{\frac{3(n_2+1)(n_2+2)}{(n+1)(n+2)(2n+3)(2n_2+3)}} \\
\times
\left[n_2\delta_{n,n_2-1}-(2n_2+3)\delta_{n,n_2}+(n_2+3)\delta_{n,n_2+1}\right]
\label{eq:fvectorTHO}
\end{multline}
(which is actually symmetric under $n \leftrightarrow n_2$, despite
appearances.) Comparing Eq.~(\ref{eq:fvectorTHO}) with the matrix
element obtained in~\cite{hys}, we get the same result for the decay
constant $f_n$ as we did in Eq.~(\ref{eq:flavorhdc}), as of course we
should.

\section{Universality}\label{sec:universality}

\subsection{$\rho$ dominance and universality}

In the limit of large $N$ and large $\lambda$, as shown in the
previous section, vector meson dominance is exact, in a sense of the
decomposition~(\ref{Ffg}). However, $\rho$ dominance cannot
be exact, on completely general grounds, at any $N$ or $\lambda$,
in a theory in which conformal invariance is exact (or violated
only by logarithmic running) in the ultraviolet.
In particular, dominance of form factors by the $\rho$ pole 
simply cannot be true in general at 
large $q^2$.  Conformal invariance in the ultraviolet
requires the form factor of
a spin-zero hadron $|a\rangle$ created by an operator of dimension $\Delta$,
must fall as $1/q^{2(\Delta-1)}$.  
More generally
\begin{equation}\label{eq:pwrlw}
\lim_{q^2\to\infty}F_{a b}(q^2)\sim
 \frac1{q^{2 k}},
\end{equation}
where $k$ depends on the spin and twist of the operator creating the
$a$ hadron. For example, $k=2$ for the form factor of the $\rho$, and
for any spin-one
hadron created by a conserved current.
This behavior, 
under the assumption that 
``vector meson dominance'' is true, requires a conspiracy
between at least $k$ poles.

Consequently the question of $\rho$ dominance
can only be relevant at small $q^2$, {\it i.e.}, the
issue is whether
\[\label{rhodom}
F_{aa}(q)\approx \frac{f_\rho g_{\rho aa}}{q^2+m_\rho^2}\ ,
\]
to some rough approximation, for small $|q^2|\lesssim m_\rho^2$.
Since $F(0)=1$, this, if true, would imply $f_\rho g_{\rho aa} \approx
m_\rho^2$, independent of $a$.  Strong $\rho$ dominance implies a
universal coupling, and sets its value.  But we will see this is not
generally true in \adscft.

However, it is logically
possible to have exactly or approximately
universal couplings without $\rho$
dominance, and in this case the universal coupling need not equal
its special value $m_\rho^2/f_\rho$.  
We will see this happens
in some sectors of \adscft.

Interestingly, the most general situation seems to
be that  $\rho$ couplings in \adscft\ contexts, though
nonuniversal, tend to lie in a rather narrow range, not varying by
more than an factor of two from $m_\rho^2/f_\rho$.  
This, combined with the structure of the spectrum,
leads to an {\it apparent} form of
$\rho$ dominance that can hold even when
the $\rho$ pole is not a dominant contributor to the form
factor at small $q^2$.  We will consider this issue in a later paper
\cite{univ}.

Let us begin the exploration of this issue with some examples that dispel any
hope of completely universal couplings.

\subsubsection*{Examples}

Here we compute $g_{n00}$ for some lowest-lying hadrons in the hard-wall and
D3/D7 model, comparing in each case the spin-one form factors for scalar and
vector hadrons, and finding they are not, in fact, universal.
Moreover, we will also find that the $\rho$ meson pole is not always
approximately dominant at small $q^2$; in fact, it is 
possible for small $n$ that $f_n g_{n00} \geq f_0g_{000}$.  (It
can even happen that
$f_n g_{naa}/m_n^2 \geq f_0g_{0aa}/ m_0^2$; we will explore this
in a later publication \cite{univ}.) 

\paragraph{\underline{D3/D7 Model}}

In the case of the D3/D7 system, the flavor form factor is easily computed
for the lowest-lying mesons within the type I scalar and the type II vector
sectors in~\ref{sec:review-models}; note the latter is the $\rho$
itself. They are
\begin{eqnarray*}
F^I_{0,0}(q)&=&\frac{6 m_h^2}{q^2+m_0^2}+
\frac{6m_h^2}{q^2+m_1^2} \to {12\over q^2} \ , \ q^2 \to \infty,\\
F^{II}_{0,0}(q)&=&\frac{12 m_h^2}{q^2+m_0^2}-\frac{12m_h^2}{q^2+m_1^2}
\to {12\over q^4}\ , \ q^2 \to\infty,
\end{eqnarray*}
where $m_n^2=4 m_h^2 (n+1) (n+2)$.   
Here the
sums over poles actually truncate, but in the first case the
truncation occurs at $n=\Delta=2$ rather than at the minimally
required $n=\Delta-1=1$.  From this simple example we immediately
learn that dominance by the $\rho$ is only
approximate even for $\Delta=2$ scalars and vectors;
in both cases the contribution of the first excited
spin-one hadron is only slightly
smaller than that of the $\rho$, since $m_1^2/m_0^2= 3$.
For the scalar, the $\rho$ contributes about
$75\%$ of $F(q^2\to0$); in particular $f_0g_{000}=\frac{3}{4} m_0^2$.
Moreover, for the scalar $\Delta=2$ hadron, where the $\rho$ 
pole {\it could} have sufficed to satisfy the power law $F(q^2)\to \#/q^2$ as 
$q^2\to\infty$, it nonetheless did not; thus we see that a 
natural guess, that conformal invariance might imply that
$f_n g_{naa}\ll f_0g_{0aa}$  for $n\geq\Delta$, is wrong, although
it happens to be correct for the form factor of the $\rho$.
Finally, universality of the $\rho$ coupling fails; since
$f_0$ is independent of the external hadron, the first terms of the two form
factors imply the corresponding $\rho$ couplings differ by a factor of two.

\paragraph{\underline{Hard-Wall Model}}

In the hard-wall model, the ground-state scalar hadron created by a
$\Delta=2$ operator has a form factor with $|f_n g_{n00}|$ peaking at
$n=2$, with $f_2 g_{200}/f_0 g_{000} \approx 5.86$.  Because $m_2^2 =
12.9\ m_0^2$, the $\rho$ and second-excited state make comparable
contributions at small $q^2$: $f_2 g_{200}/m_2^2 \approx 0.32$ and
$f_0 g_{000}/m_0^2 \approx 0.72$, while other states, including $n=1$,
make much smaller contributions, of varying sign.  Thus we again do not find strong
$\rho$-dominance, though the $\rho$ is still the most important
contribution at small $q^2$, only slightly less important than in the
D3/D7 case.  The form factor of the $\rho$ itself, on the other hand,
has an interestingly similarity to that of the $\rho$ of the D3/D7
model; $f_n g_{n 00}/f_0 g_{00}$ is $1.00$, $-1.02$ and $0.02$ for
$n=0,1,2$ respectively, with the remainder extremely small.  (We will
comment on this similarity below.)  However, the the $f_n$ and $m_n$
differ in the two models, so the $g_{n\rho\rho}$ do as well.  The
hard-wall model has $m_1^2/m_0^2 =\zeta_{1;1}^2/\zeta_{1;0}^2\approx
5.26$, compared to $4$ in the D3/D7 model, and $f_1/f_0=-3.50$ in
the hard-wall model and $\sqrt7$ in the D3/D7 model.
The deviation of $f_0 g_{000}/m_0^2$ from 1
is a bit smaller than in the D3/D7 model, about $24\%$.  Finally
the ratio of $g_{000}$ for spin-zero hadrons of $\Delta=2$ to 
$g_{000}$ (the $\rho$ self-coupling) is $0.581$, compared to
$1/2$ in the D3/D7 model.

\subsection*{}

In summary, just looking at a pair of simple states in two models, we
see both $\rho$ dominance and coupling universality violated at order
one, although not by orders of magnitude. Later we will see cases
where $\rho$ dominance is a much worse approximation, though the
$\rho$ couplings will still not vary over a large range.

\subsection{Two examples of approximate universality}

Despite the absence of $\rho$-coupling universality,
we can show that, at large 't Hooft coupling, there are two limits in which
coupling universality arises, on very general grounds.
Indeed, in these regimes the couplings of {\it all} of the
vector mesons, and indeed the entire form factor, becomes
universal, as we observed already in \cite{hys}.
Both examples stem from the simplification of mode functions in 
the associated limit. 

The first case concerns hadrons created by an operator with large
conformal dimension.  Under \adscft\ duality, an operator with
conformal dimension $\Delta$ corresponds to a five-dimensional field
whose mass is $m\approx\Delta/R$, so large $\Delta$ corresponds to a
heavy particle in five dimensions.  Gravity tends to pull particles
down to the minimal possible $\ads{}$ radius, or more precisely, to the
minimum of some effective potential due to gravity and other effects.
As always, a light particle will have a rather diffuse wave function
spread out around the minimum of this potential, while a heavy
particle will have a wave function highly concentrated at the
potential's minimum.  For example, in the duals of confining theories,
as captured in part by the hard-wall model, the normalizable mode
corresponding to a hadron created by an operator with $\Delta\gg 1$
generically is localized near the wall, where $g_{00}$ is minimized.
(This fact was used in obtaining a string theory for high-$\Delta$
hadrons in  string
backgrounds dual to a confining gauge theories~\cite{annulon}.)  In the
D3/D7 system, the normalizable modes of the flavor-charged meson-like
states localize at $\varrho=1$, where $\varrho$, which runs from 0 to infinity,
is the radial coordinate on the D7 branes introduced in
appendix~\ref{sec:review-models}.  Therefore, if we take the limit
that $\Delta\gg1$ for the hadrons $a$ and $b$,
the coupling of a vector hadron $|n\rangle$ to these
hadrons 
will only depend on the wave function $\phi_n$
of the vector hadron, evaluated at the minimum of the 
potential for the field associated to hadrons $a$ and $b$.  The
effective potential does not depend on the particle's mass, $\Delta$,
so the position of its minimum is $\Delta$-independent as well as
$a$-independent.  The
resulting overlap integral is then easily approximated
and depends only on $n$.
Consequently, because 
of the decomposition~(\ref{Ffg}), the entire
form factor $F(q^2)$ becomes independent of $a$ and $\Delta$
as $\Delta\to\infty$. 
In general, however, the convergence to the universal form factor
and couplings may be very slow.

The second case of universal couplings appears when $a=b$
and the hadron $a$ is a  very highly excited state.  In this case,
its mode function oscillates rapidly with radius. 
If the oscillation wavelength is sufficiently short, while the mode
function for  the
vector hadron $|n\rangle$ is  slowly varying, then we can
approximate the latter as constant in any region of integration,
replacing the product $|\phi_a|^2$ by its average, 
{\it i.e.}, half its maximum.
More precisely and more generally,
using the notations in
appendix~\ref{sec:methodology}, we want to compute
\begin{equation}\label{eq:gnaalargea}
g_{n a a} = g_d\int_0^{z_{max}}dz\,\mu\phi_n \phi_a^2 
\end{equation}
where $\mu= R^{5-2(S+J)}V(z)
e^{( 5-2 S - 2 J ) A ( z )}/z^{5-2(S+J)}$ 
is the metric factor and $g_d$ is the $d$-dimensional 
coupling.  (Recall $S=1$ is the spin
of the hadron $|n\rangle$
and $J$ is the spin of the hadron $|a\rangle$.) In the limit that $\phi_a(z)$
oscillates rapidly, we can use the WKB approximation,
\begin{equation}\label{eq:wkb}
\phi_a(z)\approx {\rm Re} N_a(z) e^{i \varphi_a(z)}.
\end{equation}
and
can average the oscillations to obtain
\[
g_{n a a}= g_d \int_0^{z_{max}}dz\,\mu\phi_n \phi_a^2 \approx
  \frac{g_d}{2}\int_0^{z_{max}}dz\,\phi_n(z) \mu |N_a(z)|^2.
\]
We will now show that $\mu N_a^2$ has no leading dependence on $a$, $J$, or
the conformal dimension $\Delta$ of the interpolating operator when the
excitation number $a$ gets large. Consequently, in this limit, $g_{n a
  a}$ is universal.

Let's first consider the case where $|a\rangle$ has spin zero. The mode
function $\phi_a(z)$ satisfies the Klein-Gordon equation in the
asymptotically $\ads{5}\times W$ space, which is given in
appendix~\ref{sec:methodology},
\begin{equation}
  \label{eq:kleingordon}
-\frac{1}{\sqrt{g}}(g^{zz}\sqrt{g} \phi_a(z)')'-m_a^2 g^{00}
\phi_a(z) + \tilde{m}^2\phi_a(z)=0,  
\end{equation}
where $g = (R e^{A(z)} z^{-1})^5 V(z)$, and 
$\tilde{m}$ is the {\em five dimensional} mass which corresponds
to the conformal dimension $\Delta\approx \tilde{m} R$.
Eq.~(\ref{eq:kleingordon}) can be transformed to a Schr\"{o}dinger
equation. When $\psi_a(z) = (g^{zz}\sqrt{g})^{1/2}\phi_a(z)$, we have
\[
-\psi_a''+U(z)\psi_a=m_a^2\psi_a,
\]
\begin{equation}\label{eq:potential}
U(z)=\frac{15}{4z^2}+\frac{\tilde{m}^2 R^2 e^{2A(z)}}{z^2} -
\frac{3}{16}\left[
\frac{2V'}{V} -
  \frac{(e^{2A(z)}z^{-2})'}{e^{2A(z)}z^{-2}}\right]^2 +
\frac{(V^2)''}{4V^2} 
+ \frac{3(e^{2A(z)}z^{-2}_{\bot})''}{4e^{2A(z)}z^{-2}}
\end{equation}
The approximate solution is given by
\[
\psi_a(z)\approx 
\frac{\text{const.}}{[m_a^2-U(z)]^{1/4}}
  \exp\left[i\int^z dy\, \sqrt{m_a^2-U(y)} \right]
\]
with the quantization condition
\begin{equation}\label{eq:bsquant}
\int_0^{z_{max}} dz\, \sqrt{m_a^2-U(z)} =\left(a+\half\right)\pi\quad
(a=0,1,2,\ldots).
\end{equation}
While $U(z)$ is fixed by the metric and $\tilde{m}$, the mass $m_a$ can be
arbitrarily large as we increase $a$. Thus, we can take the limit that
$m_a$ is so large that $U(z)$ is negligible except at small values of 
$z$ (where the contribution to the $g_{naa}$ integral is small.)
In this limit,
$\sqrt{m_a^2-U(z)}\approx m_a$, and we obtain $m_a \approx \pi
a/z_{max}$ from the quantization~(\ref{eq:bsquant}). Also, in this
limit, $\psi_a(z)$ can be further approximated as
\[
\psi_a(z)\approx \tilde{N} \exp(im_a z),
\]
where $\tilde{N}$ is a constant which can be determined by the
normalization condition~(\ref{eq:normal}). Therefore, $N_a(z)$ and
$\varphi_a(z)$ in Eq.~(\ref{eq:wkb}) are completely fixed:
$N_a(z)\approx \tilde{N}(g^{zz}\sqrt{g})^{-1/2}$, which is
independent of $a$ and $\Delta$, and $\varphi_a(z)\approx m_a z$.
From this it follows that $g_{naa}$ is independent of $a$
and $\Delta$.

The approximation $\sqrt{m_a^2-U(z)}\approx m_a$ breaks down when
$U(z)$ becomes of order $m_a$, which generally
occurs in the small-$z$ region.
Here the term in $U(z)$ depending on the conformal dimension in
Eq.~(\ref{eq:potential}), $15/4z^2+\tilde{m}^2 R^2 e^{2 A(z)}/z^2
\approx (4 \Delta^2 e^{2 A(z)}+15)/4z^2$, diverges. Since the space is
nearly $\ads{5}\times W$ in the region, this term can be neglected for
\[
\frac{\Delta^2+\frac{15}{4}}{z^2}\ll m_a^2\quad \Leftrightarrow\quad
{z}\gg\frac{\sqrt{\Delta^2+\frac{15}{4}}}{a}z_{max}.
\]
In other words, the region where our approximation is not valid will
expand as $\Delta$ gets large with $a$ fixed, and this calculation
is valid only for $\Delta\ll a$.  But in the small $z$ region,
the wave function matches on to a known $z^{-\Delta}$ power law,
and so the contribution of the small $z$ region to any calculation
is generally small, especially at large $\Delta$.

Our discussion so far can be easily generalized to the case where
$|a\rangle$ has spin $J$.  In the WKB approximation~(\ref{eq:wkb}),
$\varphi_a\approx m_a z$ and
\begin{equation}\label{eq:largennormal}
N_a(z) \approx (z/R)^{(3-2J)/2}
\frac{e^{(J-3/2)A(z)}}{\sqrt{z_{max}V(z)}
}.
\end{equation}
and consequently
\[
\mu
N_a(z)^2 \approx (R e^{A(z)}/z)^{2(1-S)}
\]
Thus the integrand in (\ref{eq:gnaalargea}) is independent of $a$,
$\Delta$ and $J$ for $a\gg \Delta$.  Consequently, as before,
$g_{naa}^\Delta$ depends only on $n$ in this limit, and so
all hadrons of any $J$ with $a\gg \Delta$ have
a universal form factor.

\subsubsection*{Examples}

\paragraph{\underline{Large Dimension}}

First we consider the case of $\Delta\gg a$, or in the notation
that we have used for the examples, $\Delta\gg n_1,n_2$.
In the hard-wall model, as we mentioned earlier, a normalizable
mode associated to a hadron $|n_1\rangle$
localizes at the wall, $z=z_{max}=1/\Lambda$. Thus, for
$n\ll n_1,n_2\ll \Delta$, the first kind of
universal vector hadron coupling is given by the value of the
normalizable mode $\phi_n(z_{max})$:
\begin{equation}\label{eq:cunivhard}
g^\Delta_{n,n_1,n_2} \underset{\Delta\to\infty}{\longrightarrow}
     \delta_{n_1,n_2}\frac{(2\pi)\sqrt2}{N},
\end{equation}
where we denoted the conformal dimension of the operator creating the
other two hadrons by $\Delta$.  Note $g^\Delta_{n,n_1,n_2}\sim 1/N$ is
consistent with $N$-counting analysis.  
That this is $\Delta$-- and $n_1$--independent
is as we expected.  That it is $n$-independent appears to be
an accident of the hard-wall model, in particular, a special
property of the Bessel equation; we do 
not expect this to hold in general 
models. On the other hand, the fact that $g^\infty_{0,n_1,n_1}$
is {\it nonzero} is generically the case, since Neumann boundary
conditions are required for a conserved current, making $\phi_0(z_{max})=0$
unlikely and indeed unnatural.
Note, however, that the limit in which \Eref{eq:cunivhard} applies is
attained only very slowly as $\Delta\to\infty$.

\FIGURE[ht]{
\epsfbox{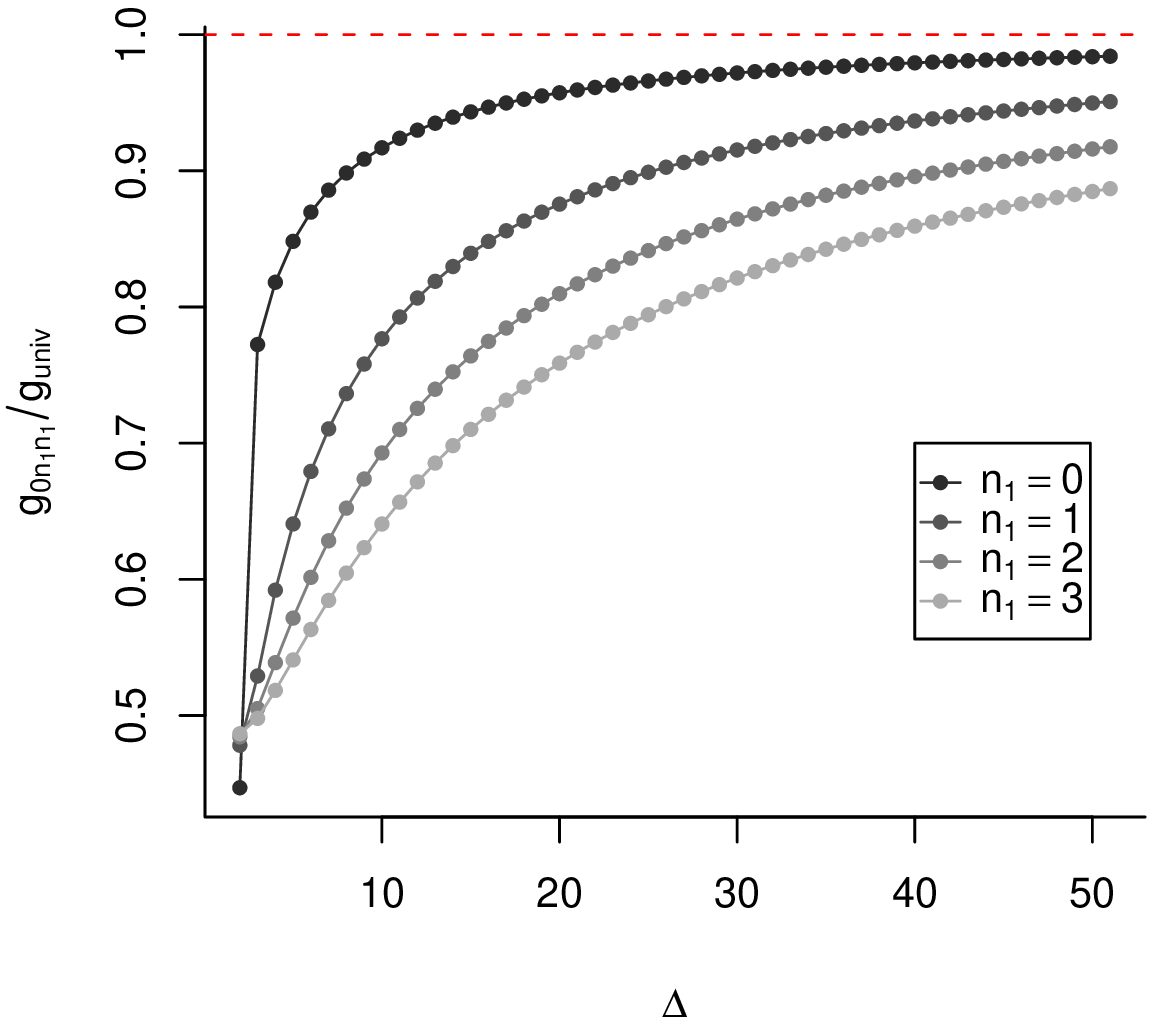}
\caption{
In the hard-wall model, as a function of $\Delta$,
the ratio of the coupling $g_{0,n_1,n_1}^\Delta$ for scalar hadrons
(showing curves for $n_1=0,1,2,3$)
and its universal value $g_{univ}$,
given in Eq.~(\ref{eq:cunivhard}).
}
\label{fig:largel}
}

In the D3/D7 system, the vector meson coupling
to two other mesons of any kind has a universal limit,
\begin{equation}\label{eq:funiv}
  g^\Delta_{n,n_1,n_2} \underset{\Delta \to\infty}{\longrightarrow}
      \delta_{n_1,n_2} \frac{(2\pi)}{2 \sqrt{N}}C^{II}_{0,n}P_n^{(1,1)}(0) =
      \delta_{n_1,n_2} \frac{(2\pi)}{\sqrt{2N}}\sqrt{\frac{(2n+3)(n+2)}{n+1}}P_n^{(1,1)}(0).
\end{equation}
Again, $g^\Delta_{n,n_1,n_2}\sim 1/\sqrt{N}$ is consistent with
$N$-counting. In this case, we can compare this value with other
explicit computations. In~\cite{hys}, $g_{n,0,0}$'s were computed
for some specific cases; it can be checked that they have the same
limit in the large conformal dimension, which, as we have just
argued, is no coincidence.  Indeed, using 
\[ P^{(1,1)}_n(0) =
\frac{(n+1)}{2^{n}}F(-n,-n-1;2;-1) =
\frac{2\cos\frac{n\pi}{2}}{\sqrt{\pi}}\frac{\Gamma\left(\frac{n}{2} +
    \frac{3}{2}\right)}{\Gamma\left(\frac{n}{2} + 2\right)}
\]
we see
that Eq.~(\ref{eq:funiv}) exactly matches with the limit
of Eq.~(5.7) in~\cite{hys}.
Note that $\lim_{\Delta\to\infty}
g^\Delta_{n,n_1,n_2}$ vanishes when $n$ is odd. When $n=2j$ is even,
\begin{equation}\label{eq:gDelta37}
\lim_{\Delta\to\infty} g^\Delta_{n,n_1,n_2}
\underset{n\to\infty}{\longrightarrow} 
\delta_{n_1,n_2} (-1)^j\frac{(2 \pi)}{\sqrt{N}} {2\sqrt2 \over \sqrt{\pi}},
\end{equation}
whose magnitude is $n$-independent, and differs by only ten percent from
\begin{equation}\label{eq:grhoDelta37}
\lim_{\Delta\to\infty} g^\Delta_{0,n_1,n_2}
= 
\delta_{n_1,n_2}\frac{(2 \pi)}{\sqrt{N}} \sqrt{3},
\end{equation}

We have mentioned that the form factors satisfy the power law
Eq.~(\ref{eq:pwrlw}) due to the conformal invariance of the field
theory in the ultraviolet. In the large conformal dimension case, the
power $k$ in Eq.~(\ref{eq:pwrlw}) diverges, so we would expect that in
our present approximation this would appear as exponential fall-off at
large $q^2$. In the hard-wall case, it is not immediately obvious.
Given the universal couplings that we have just observed, it would
seem that expansion of the form factor as a sum of poles,
Eq.~(\ref{Ffg}), diverges: the coefficient $f_n g^{\infty}_{n} \sim
{\cal O}(n^{3/2})$, where $g^{\infty}_n \equiv
\lim_{\Delta\to\infty}g^{\Delta}_{n a a}$ is the universal coupling.
It is possible to compute the sum by regularizing it carefully, or
redefining it by one or more subtractions, but instead we can easily
evade the problem altogether. Recalling that the form factor is
obtained by the same overlap integral as the tri-meson coupling,
Eq.~(\ref{eq:formfactor}), but with the normalizable mode of the mediating
vector meson replaced by a {\it non-normalizable mode}, we can apply
the same approximation to the integral as we did for
$g_{naa}^\Delta$. In particular, the large-$\Delta$ hadrons of the
hard-wall model will have a universal form factor given by the value
of the non-normalizable mode at the wall. From Eq.~(\ref{eq:Ahadnorm}), 
we use the identity $K_n(x) I_{n+1}(x) +
K_{n+1}(x) I_{n} (x) = 1/x$ at $x = q z_{max}=q/\Lambda$ to obtain
\begin{equation}\label{eq:univFF }
 F_{ab}(q)= \delta_{ab} {1\over I_0(q/\Lambda)}.
\end{equation}
Indeed it vanishes exponentially at large spacelike $q^2$, as we expected.

Similarly, in the D3/D7 case, the universal form factor of the flavor
current is given by
\begin{equation}
F_{ab}(\q)=
-{2\pi^{3/2} \over \sin(\pi\alpha)\Gamma\left(-\frac\alpha2\right) \Gamma\left(\frac{1+\alpha}2\right)}
%
\end{equation}
where $\q=q/m_h$ and $\alpha=(-1+\sqrt{1-\q^2})/2$.  This too falls
off exponentially at spacelike $q^2$.

\paragraph{\underline{Highly Excited Hadrons}}

Now let's turn to the universality for highly excited hadrons,
$a\gg \Delta$.
In the hard-wall
model, any spin-$J$ normalizable modes, such as
Eqs.~(\ref{Ahadwave}) and~(\ref{eq:hhadwave}), can be approximated as
\[
\phi^{(\Delta)}_{p;n}(z)\sim z^{2-J} J_{p} (\zeta_{p-1;n}\Lambda z ) \approx
\sqrt{\frac{2}{\pi}} z^{3/2-J}\sin (\zeta_{p-1;n}\Lambda z ),
\]
for large $\zeta_{p-1;n}\Lambda z$, where $p$ is a constant 
depending on $\Delta$ and
$J$, and
$\zeta_{p-1;n}$ is the $n$-th zero of the Bessel function $J_{p-1}(x)$.
The three hadron coupling for a vector hadron $|n\rangle$
and a spin-$J$ hadron $|n_1\rangle$ is
\[
g^{\Delta}_{n,n_1,n_1}=
\frac{\sqrt{2}\kappa \Lambda}{R^4} R^{8-2J} \pi^3 \int^{1/\Lambda}_0
\frac{dz}{z^{3-2J}} \, \frac{zJ_1 ( \zeta_{0 ; n}\Lambda z )}{\pi^{\frac32}
J_1 (\zeta_{0 ; n} )} |\phi_{p,n_1}(z)|^2 
\]
Since for small $z$ the mode functions are all
power-law suppressed, the integral can be approximated
using the sine-wave form for the external
hadrons, giving
the approximately universal coupling
\begin{eqnarray*}
g^\Delta_{n,n_1,n_1} & = &\frac{(2\pi) 2^{\frac32}}{N} \int^1_0 d\hat z \,
  \frac{\hat zJ_1 ( \zeta_{0; n}\hat z )}{J_1 ( \zeta_{0 ; n} )} |\sin ( m_{p;
  n_1} \hat z/\Lambda )|^2
\\
  & \approx & 
\frac{(2\pi) \sqrt2}{N} \int^1_0 d\hat z \,
  \frac{\hat zJ_1 ( \zeta_{0 ; n} \hat z )}{J_1 ( \zeta_{0 ; n} )}\\
  & = & 
\frac{(2\pi)\sqrt2 }{N}\cdot\frac{\pi}{2\zeta_{0;n}}
  \mathbf{H}_0(\zeta_{0;n}),\label{eq:THOlargen-hw}
\end{eqnarray*}
where $\mathbf{H}_\gamma$ is the Struve-H function.  In particular,
\begin{equation}
g^\Delta_{0,n_1,n_1} 
\approx
0.490\ \frac{(2\pi)\sqrt2}{N}
   \
.
\end{equation}

It is interesting to compare this last result with
Eq.~(\ref{eq:cunivhard}).  The large-$\Delta$ and large-$n_1$ limits
do not have the same $\rho$ couplings, and thus the two limits do not
commute.  However, the couplings in these limits differ only by a
factor of about two.  Moreover, $g^{\Delta=2}_{0,0,0} = 0.447\ 
\frac{(2\pi)\sqrt{2}}{N}$ for spin-zero hadrons, also quite close to
both limits.  Indeed, we seem to find that, over the whole domain of
$\Delta$ and $n_i$, the couplings of the $\rho$ vary within a rather
narrow range.  There is no exact universality in this model, but we
see no drastic violation of it either.

\FIGURE[ht]{
\epsfbox{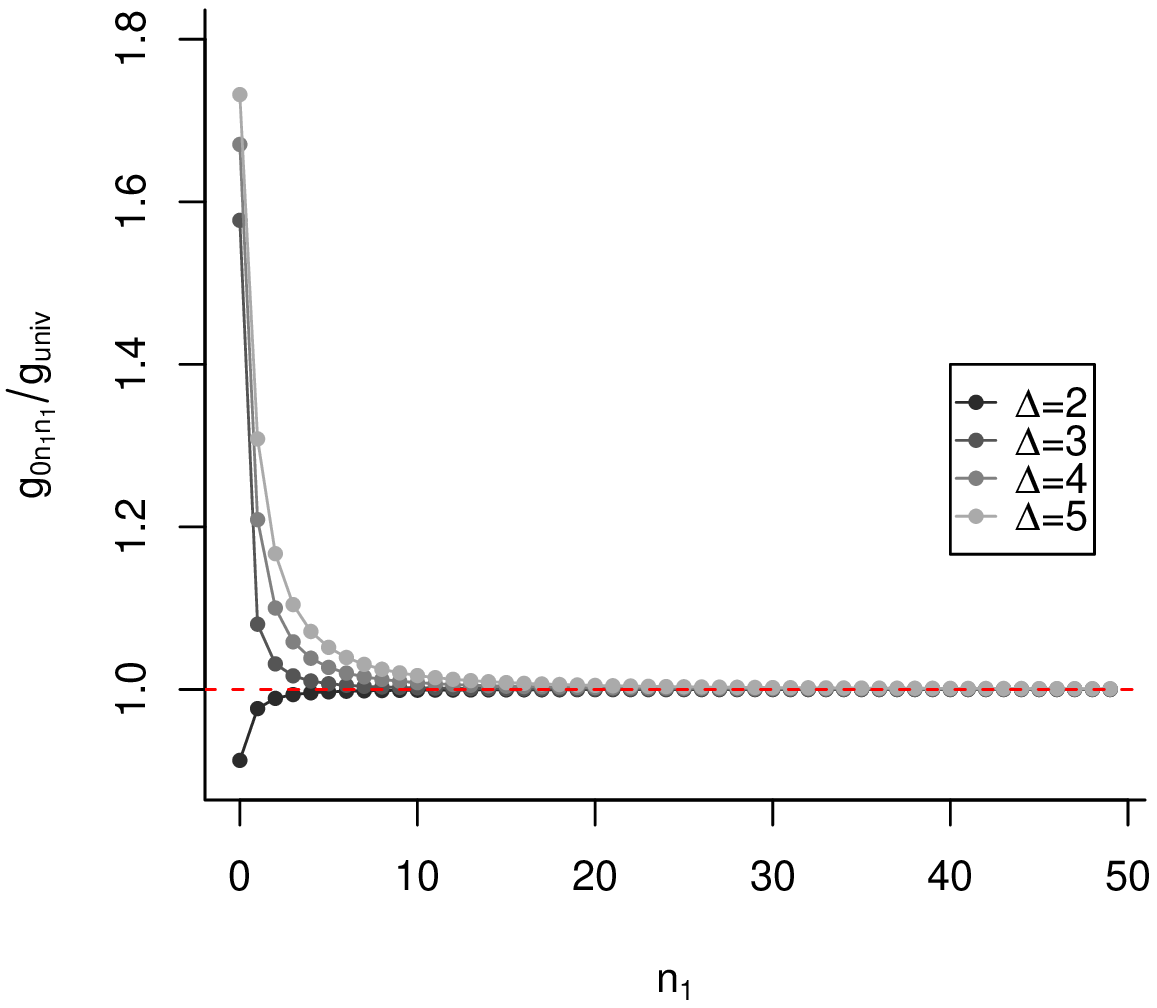}
\caption{
In the hard-wall model, as a function of $n_1$,
the ratio of the coupling $g_{0,n_1,n_1}^\Delta$ for scalar hadrons
and its universal value $g_{univ}$,
given in Eq.~(\ref{eq:THOlargen-hw}).
}
\label{fig:largen}}

In the D3/D7 system, we use a similar approximation 
\begin{equation}\label{eq:jacapprox}
P^{( \alpha, \beta )}_n ( 2 v - 1 ) = \frac{\cos \left\{ \left[ 2 n +
      ( \alpha + \beta + 1 ) \right] \cos^{- 1} {v}^{1/2} - \left(
      \frac{1}{2} \alpha + \frac{1}{4} \pi \right) \right\}}{\sqrt{\pi
    n} ( 1 - v )^{\alpha/2 + 1 / 4} v^{\beta/2 + 1 / 4}} +\mathcal{O}(
n^{- 3 / 2} ).
\end{equation}
This leads to a similar approximation of the overlap integral,
\begin{equation} \label{eq:THOlargen}
g^\ell_{n,n_1,n_1} \approx g_8 \frac{{2}}{\pi} \int_0^\pi
d\theta\, \phi_n(\theta) |\cos n_1 \theta|^2 
\approx
\frac{\sqrt{2} (2 \pi)}{\sqrt N}R^2\int_0^\pi d\theta\,
\phi_n(\theta)
\end{equation}
where $\cos\theta=2v-1$. We can check that
Eq.~(\ref{eq:THOlargen}) matches with the form factor computation
(5.8) in~\cite{hys},
though this is rather trivial since both results are
derived from the same approximation~(\ref{eq:jacapprox}).
The result is
\begin{eqnarray}
g^\ell_{n,n_1,n_1} &\approx&
\frac{(2\pi) \sqrt{2}}{\sqrt{N}} 
\sqrt{\frac{(2n+3)(n+2)}{\pi(n+1)}} \nonumber \\
& & \quad \times (-1)^n\frac{(n-\half)!}{n!}{{}_3F_2}\left(\frac32,-n,-n-1;2,\half-n;1\right).
\label{eq:THOlargenB}
\end{eqnarray}
This expression grows as $\sqrt{n}$ for $1\ll n\ll n_1$.
For the $\rho$, the $n=0$ case, we have
\begin{equation}\label{eq:THOlargenrhoB}
g^\ell_{0,n_1,n_1} \approx 
\frac{(2\pi)}{\sqrt{N}} 
\sqrt{3} \ .
\end{equation}

Note that Eqs.~(\ref{eq:gDelta37}) and (\ref{eq:THOlargenB}) differ,
although for the $\rho$ meson, interestingly, the large-$\Delta$ and
large-$n_1$ limits give the same result, Eqs.~\eref{eq:grhoDelta37} and
\eref{eq:THOlargenrhoB}.   We will see this can be viewed as resulting from
the exact universality that we discuss in the next section. 
Also, \eref{eq:THOlargenrhoB}
is of the same order as 
the couplings of the $\rho$ to the lowest-lying mesons in the theory.  
In particular, $g^{\ell=1}_{000}$ for external scalar
hadrons is given in Eq.~(\ref{eq:scalarTHO}), $g^{\ell=1}_{000} =
\frac{(2\pi)}{\sqrt{N}}{\sqrt{3}\over 2}$, while $g^{\ell=0}_{000}$ for
external vector hadrons, from Eq.~(\ref{eq:fvectorTHO}), is 
again $g^{\ell=0}_{000} = \frac{(2\pi)}{\sqrt{N}}\sqrt{3}$.

As was the case for large $\Delta$, the universal
form of the couplings implies a universal
form factor.
Applying our approximation
strategy for high excitation modes to the form factor calculation, we find
a universal form factor for highly excited hadrons in the hard-wall
model,
\begin{equation}
\lim_{n_1\to\infty} F_{n_1,n_1}(q)=
\frac{\pi}{2}\mathbf{L}_0(q/\Lambda)
\left[K_1(q/\Lambda) + \frac{K_0(q/\Lambda)}{I_0(q/\Lambda)}I_1(q/\Lambda)\right]
\end{equation}
where $\mathbf{L_\gamma}$ is the Struve-L function. Similarly, highly
excited hadrons in the D3/D7 model have the universal form factor
\begin{equation}
\lim_{n_1\to\infty} F_{n_1,n_1}(q) =
\frac{4}{\q^2} -
\frac{\pi}{\cos\left(\frac{\pi}{2} \sqrt{1-\q^2}\right)} 
\end{equation}
where $\alpha = (-1+\sqrt{1-\q^2})/2$.

\paragraph{\underline{Additional Comments}}

In some of the above examples,
the $g_{naa}$ seem to exhibit $n$-independence, 
or even growth with $n$, for large $n$.  This behavior must
break down, because of the power law~(\ref{eq:pwrlw}).
From~(\ref{Ffg}), the power law can only hold if the
moments $\sum f_n g_{n a b} m_n^{2j}$, for all $j<k$ [where $k$ is
the power in Eq.~(\ref{eq:pwrlw}),] vanish.  Truly universal
$a$-independent and/or $\Delta$-independent couplings
$g^\Delta_{naa}$, for high but fixed $a,\Delta$ and for all $n$, would
endanger this power law.  Consequently, any universality with respect
to $n$ must break down eventually.  If $n$, $a$ and $b$ are all very
large, the computation of $g_{nab}$ involves the overlap integration
of a product of three rapidly oscillating functions, and for
sufficiently large $n$ this will begin to decrease.  Similarly, when
$n\gg\Delta$ the spin-one mode $\phi_n$
oscillates so quickly that one cannot treat the
external hadron as localized on the scale of the oscillations.

An interesting pattern which appears in both models concerns the
couplings of the $\rho$.  Along with all the other vector meson
couplings, $g_{\rho aa}^\Delta$ has a universal value at large
$\Delta$, and a second universal value at large $a$.  These differ,
but are of the same order, in the hard-wall model; in the D3/D7 model
the two limits commute, for reasons that we will see in the next
section.  In all cases the coupling in these limits differ from the
$\rho$-dominance prediction $m_\rho^2/f_\rho$, given in
Eqs.~\eref{eq:grhohardwall} and \eref{eq:grhod3d7}, but only by a
factor of order two. Finally, neither differs much from the
(non-universal) couplings of the $\rho$ to the lowest-lying mesons in
the theory, including its own self-coupling.  In short, we do not find
that the conjecture of universal couplings is true, but neither do we
find it badly violated.  This deserves an explanation, which none of
the arguments presented in this paper directly provides.  We will
address this issue further in a future publication \cite{univ}.

\subsection{Exact universality}

Amusingly, we have found one example of exact universality for the
couplings of the $\rho$ to a certain class of hadrons.  As hinted
already by some of our earlier calculations, this arises in a
subsector of the D3/D7 system.  The universality can be derived from a
certain symmetry satisfied by the relevant mode functions, but we have
not found that this mathematical property of the modes has any deeper
physical significance.  A similar sector in the hard-wall model does
not show exact universality.  At this level, then, the example we now
present appears special to this model, and in this sense, accidental.

We begin with the type II modes, of which the $\rho$ is one,
which lie within the subsector exhibiting universality.
The coupling of three type II modes is computed using~(\ref{eq:FFF}). One
portion of the integrand involves two mode functions and
a metric factor
\begin{equation}\label{eq:prodmodeII}
  \left( \frac{1-v}{v} \right)\phi^{II}_{\ell,n_1} \phi^{II}_{\ell,n_2}
= R^{-4} {\hat C}^{II}_{\ell n_1} {\hat C}^{II}_{\ell n_2} v^{\ell+1}
(1-v)^{\ell+1}P^{(\ell+1,\ell+1)}_{n_1}(2 v -1)
P^{(\ell+1,\ell+1)}_{n_2}(2 v -1) \ .
\end{equation}
This is invariant under the transformation $v\to 1-v$, up to the sign
$(-1)^{n_1+n_2}$.  The remaining factor of the integrand is the wave
function $\phi^{II}_{0,n}$, which transforms non-trivially under $v
\to 1-v$, or equivalently $\varrho\to 1/\varrho$.  We can decompose
this function into the sum of odd and even parts. The higher is $n$,
the more complicated is each part, but for the $\rho$ meson, the
lowest mode $n=0$, the wave function is very simple:
\[
\phi^{II}_{0,0} = R^{-2} {\hat C}^{II}_{0, 0} v =
R^{-2} {\hat C}^{II}_{0, 0} \left[ \half + \left(v-\half\right)\right].
\]
If $n_1-n_2$ is even, then the odd part of this function can
be dropped; the even part is constant and the
computation reduces to the normalization integral of the modes.
Therefore, we find that within this sector the $\rho$ has
a universal diagonal coupling
\begin{equation}\label{eq:d3d7coupling}
g^\ell_{\rho,n_1,n_1}=
\frac{(2 \pi)}{\sqrt{N}}\sqrt{3}.
\end{equation}

This result extends beyond the type II modes, due to an accidental
symmetry relating the type II mode to others. It has been found that
the D7 brane worldvolume theory has extra degeneracy among the scalar,
type II and type III modes.  This degeneracy is not required by any
obvious symmetry, but its presence has been interpreted as a sign of
an extension to $SO(5)$ of the explicit $SO(4)$ present in the
classical field theory~\cite{Myers}.  We will refer to the degenerate
modes as the ``$SO(5)$ multiplet.'' Note that the $SO(5)$ symmetry
relates states with different spin. For example, it relates the
pion-like scalar mesons created by ${\psi_Q}^\dag\Phi^\ell\psi_Q$, and
the $\rho$-like spin-one mesons. All the modes in the $SO(5)$
multiplet are eigenvectors of the transformation $v \to 1-v$, which as
we have just seen above leads them to have the universal coupling
(\ref{eq:d3d7coupling}) to the $\rho$, and indeed, to all of the
ground states related to the $\rho$ by $SO(5)$.

It is easy to check that the scalar and the type III modes have the
same behavior and universal coupling as those of type II. For the
scalar mode, the product of the wavefunctions and the metric factor
gives exactly Eq.~(\ref{eq:prodmodeII}), just as for type II,
so the same
universality is trivially obtained. The type III modes are
different in appearance, as they
involve gauge fields polarized both in
the compact $S^3$ directions and the fifth dimension, but in the 
end the integral is the also the same as for type II.

~From \eref{eq:d3d7coupling}, it follows that the large dimension limit
$\ell\to\infty$ and the large excitation limit $n_1\to\infty$ lead to
the same limiting $\rho$ coupling in these particular sectors.  But the
coupling arising in each limit is the same in {\it all} sectors.
Therefore, in all sectors, the $\rho$ coupling must approach
\eref{eq:d3d7coupling} both at large dimension and at large
excitation.  This explains why
Eqs.~\eref{eq:grhoDelta37} and \eref{eq:THOlargenrhoB} agree
with each other and with \eref{eq:d3d7coupling}.

As we noted, the key fact leading to universal couplings 
is that the integrands in Eqs.~(\ref{eq:normal}) and
(\ref{eq:THO}) are identical except for the mode function $\phi_n$
of the spin-one vector meson.
One might ask if there are other natural contexts where symmetries
might constrain $\phi_n$, or in particular the $\rho$ mode function
$\phi_0$, such that the overlap computation would reduce to the
normalization integral~(\ref{eq:normal}), giving a universal value for
the $\rho$'s coupling to all hadron states. In the D3/D7 case above,
the ``parity'' $v \to 1-v$ (really an inversion symmetry $\varrho\to
1/\varrho$) played such a role.  It would be interesting to build this
feature into a model to obtain the universality seen in QCD, something
along the phenomenological lines of \cite{SonSteph}.  We
leave this question for
future study.

\subsubsection*{Examples}

We already have computed in section~\ref{sec:decomp} one example of
a tri-meson coupling in the type II sector,
Eq.~(\ref{eq:fvectorTHO}). Letting $n=n_2$, and using the cyclic symmetry of
the coupling~(\ref{eq:FFF}), $g^{\ell}_{0,n,n}=g_{n,0,n}$,
we see that $g^0_{0,0,0}$ is indeed the exactly
universal coupling $\frac{(2 \pi)}{\sqrt N} {\sqrt3}$.  Note however
that $m_\rho^2/f_\rho$, Eq.~\eref{eq:grhod3d7}, 
is smaller by a factor of $\frac{3}{2}$.

\section{Conclusions and discussion}\label{sec:concl-disc}

We have examined the universality of the $\rho$ meson's couplings, and
those of excited vector mesons, in the \adscft\ context.  We did not
find that the $\rho$ typically has precisely universal couplings.  We
did find two regimes of approximate coupling-universality, which
become exact in certain limits.  These are especially interesting
because they are generic, arising for fundamental reasons which apply
in any theory at large 't Hooft coupling.  The first case is the
$\rho$'s couplings (and those of other vector mesons) to hadrons
created by interpolating operators of very large conformal dimension;
in this case, universality stems from the localization of the
associated mode functions at the minimum of an appropriate effective
potential.  The second case involves hadrons which are highly excited
states; the wave functions oscillate rapidly, and these fluctuations
average out in the calculation of the couplings to vector mesons.
Note that in general the two different universalities do not commute
with each other, indicating that they represent two distinct regimes.
Moreover, as we saw examining two models, the $\rho$'s couplings do
not match the conjectured value $m_\rho^2/f_\rho$ in either regime,
though they do not differ from it by more than a factor of two in
either model.

We also saw that a large sector of one model (the D3/D7 system)
exhibits exact coupling-universality for the $\rho$.  As a
consequence, the two above-mentioned limits commute in this model.
This feature requires special properties which constrain the mode
function of the $\rho$ relative to the other modes.  We expect this
behavior is highly model-dependent and does not generically arise
elsewhere.

For further study, then, there are two main questions that we should
ask.  First, why are the $\rho$ couplings often roughly universal, and over
what range can they vary in generic models?  Clearly there is a
connection with the fact that the $\rho$ is created by a conserved
current, which has the special property that its non-normalizable mode
at $q^2=0$ is always a constant in the radial direction, with a fixed
normalization, in order to ensure $F(q^2\to 0)\to 1$.  Second,
why does $f_\rho g_{\rho aa}$ tend to be of order $m_\rho^2$ {\it even
when} many other vector mesons are large contributors to
a form factor?  This, too, is presumably tied to the particular
shape of the $\rho$ meson's mode function, which, being generally
positive definite and structureless, is significantly constrained.
We leave these questions for further study \cite{univ}.

There are several other interesting problems which were not dealt with
in this paper. One is the computation of corrections beyond
supergravity.  There have been many interesting approaches to this
problem \cite{DIS,rho,janik,bmn,tseytlin}.  In particular, it has been
noted that the string theory can be highly simplified for large
conformal dimension $\Delta\sim\sqrt{N}$~\cite{bmn}, where one of our
examples of universality arises.  Aided by these simplifications, this
limit may serve as a nice testing ground to discover more interesting
relationships between string theory and QCD.

The conjecture of universal couplings includes nucleons as well.  We did
not consider baryons here, as at large $N$ they are very different
objects from mesons.  Indeed, at large 't Hooft coupling they are
described by D-branes rather than of supergravity modes.  At present
there is no suitable tool for the relevant computations; the baryons'
charges can be calculated using geometry, but the formalism for
computing dynamic quantities such as a form factor is still undeveloped.
Still, the baryons are localized at small radius in much the same way
as $\Delta\gg1$ mesons, and for much the same reason.  We might
therefore expect that they share the same universal form factor as
large-$\Delta$ mesons, but this remains to be confirmed.  For
five-dimensional states with extremely large mass
(corresponding to
field-theoretic operators of dimension much larger than $N$) 
back-reaction on the metric eventually
becomes important.  \cut{Indeed, there have been
interesting speculations about production of ultra heavy-objects such
as black holes, etc.\cite{giddings,arkanihamed}.}  This back-reaction
would be relevant, for instance, for nuclei with large numbers of
baryons.  Although these
objects, too, tend to localize due to their heavy mass, their
backreaction on the metric is likely to alter their form factors.

One may consider the experimental implications of these
findings, but a little thought reveals the situation is not
encouraging for any direct application. It is very difficult to
measure tri-vector-meson couplings, even $g_{\rho \rho \rho}$, or form
factors of unstable particles, even the $\rho$.  As we have discussed
and have seen in the examples, there is no reason to believe that
$g_{\rho\rho\rho}$ is approximately equal to, for example, $g_{\rho
\pi \pi}$, at least in the large $\lambda$ limit.  Indeed our examples
suggest that $g_{\rho\rho\rho}$ can differ from $g_{\rho\pi\pi}$ by a
factor of 2 or so.  An attempt could be made to measure
$g_{\rho\rho\rho}$ in the process $\pi^+ p \to \pi^+ \rho^+ n$, but to
extract $g_{\rho\rho\rho}$ in a fully model-independent way would not
be possible in this experiment; one would have to assume $\rho$
dominance in the intermediate states.  Meanwhile, the approximate
universalities that we found for certain states in the \adscft\
models are completely out of experimental reach.
Perhaps there are more subtle ways to apply our results to QCD, but we
will have to seek them in the future.

Still, it is interesting to observe that although most of our results
are, in a sense, negative, in that we do not confirm the classic
conjectures, we still have the unexplained fact that the $\rho$
couplings to most objects in the theory appears to be of the same
order.  The structure of the calculation in \adscft\ seems to suggest
that this arises from profound properties of mesons created by
conserved currents. In this sense, Sakurai's original idea of treating
the $\rho$ as a gauge boson seems not entirely misguided.  We will
return to this issue in \cite{univ}.

\acknowledgments We thank Joshua Erlich, David Gross, Andreas Karch,
Ami Katz, Mikhail Stephanov, Martin Savage and Dam Thanh Son for
useful conversations.  This work was supported by U.S. Department of
Energy grants DE-FG02-96ER40956 and DOE-FG02-95ER40893, and by an
award from the Alfred P. Sloan Foundation.

\appendix

\section{Review of methodology}\label{sec:methodology}

The methodology we use in this paper is established originally in the
hard wall model~\cite{hardscat,polchinskisusskind,DIS} and applied to
the the D3/D7 system~\cite{hys}.

We first assume that our confining model is given by the
asymptotically $d$-dimensional $\ads{5}\times W$ space ($W$ a
compact manifold of dimension $d-5$) which has
the following metric
\begin{equation}\label{eq:metric}
d s^2 =
\frac{R^2  e^{2 A ( z )}}{z^2} ( \eta_{\mu\nu}dx^\mu dx^\nu +
  dz^2)
+\hat{g}_{\bot i j} d \hat{z}^i d \hat{z}^j \ ,
\end{equation}
where $e^{2 A(z)}\to 1$ as $z\to 0$. $x^\mu$ is tangential to the four
dimensions, $z$ the $\ads{}$ radius, and $\hat{z}^i$'s are the
coordinates on $W$.  $\hat{g}_{\bot i j}$ is the metric of $W$.  We
assume that the square of the warp factor $e^{2 A ( z )}/{z^2}$ has a
minimum at $z=z_{max}$. This is one of the sufficient conditions that
this background is dual to a confining gauge theory~\cite{sonn18}.
Also, in the light-cone gauge, the warp factor squared has a natural
interpretation as the potential for a classical long
string~\cite{polchinskisusskind} in the string action. Therefore,
$z_{max}$ can be interpreted as ``the wall at the end of
space''~\cite{sonn18}, beyond which a string cannot go.

Each hadron state is dual to a {\em normalizable} mode in five
dimensions.  When a spin-$J$ ($J\leq2$) field is given by
$C_{\mu_1\mu_2\ldots\mu_J}=\epsilon_{\mu_1 \mu_2\ldots\mu_J}
e^{ik\cdot x} \phi(z) \ylm(W)$, then the normalizable mode $\phi_n(z)$ with
$k^2=-m_n^2$ satisfies the normalization condition
\begin{equation}
  \label{eq:normal}
R^{3-2J}\int_0^{z_{max}} \frac{dz}{z^{3-2J}}\,
  e^{(3-2J)A(z)} 
V(z)
  \phi_{n_1} \phi_{n_2}=\delta_{n_1n_2},
\end{equation}
where $V(z)$ is a normalization coefficient in $W$
direction
\begin{equation}
\label{eq:Vdef}
V(z) = \int d^{d-5}\hat{z}\, \sqrt{g_\bot} \left| \ylm(W) \right|^2.
\end{equation}
In principle, we might encounter hadron states dual to bulk vector
or rank-two tensor fields which are partially or entirely polarized in
the $\hat{z}^i$ directions. In such cases, we would have to include
suitable $\hat{g}^{ij}$ factors in the integrand of Eq.~(\ref{eq:normal}).
However, we can absorb such factors into the wavefunctions, and treat
such fields as five dimensional scalar or vector fields
satisfying Eq.~(\ref{eq:normal}). 
Also note that $V(z)$ depends on the normalization of
  $\ylm(W)$ on $W$, which can be arbitrarily chosen. In this paper,
  we use the convention that the norm of $\ylm(W)$ is equal to the
  volume of $W$. In particular, this sets the lowest constant mode
  ${\cal Y}^0(W) = 1$, and $V(z)$ is the volume of $W$ at $z$.

To compute the matrix element of a current, we need the
{\em non-normalizable} mode dual to that current. Then we find the
trilinear interaction between the three modes corresponding to the
initial state, the final state and the current operator. Such an
interaction can be derived either from bulk supergravity or the
Born-Infeld action on D7 branes if present. The matrix element for the
spin-$S$ current and the spin-$J$ hadrons is given by 
\begin{equation}
\bra{b}\mathcal{J}^{\mu_1\mu_2\ldots\mu_S}\ket{a} = ( \text{charge} )
  \times ( \text{kinematic factor}  ) \times F_{a b} ( q^2 ),
\end{equation}
where the form factor $F_{a b} ( q^2 )$ is\footnote{Here we again ignore
that we need multiple form factors depending on $J$ and whether
$\mathcal{J}$ is conserved as discussed in~\ref{sec:decomp}.}
\begin{equation}
  \label{eq:formfactor}
F_{a b} ( q^2 ) = g_d R^{5-2(S+J)}\int \frac{dz}{z^{5-2(S+J)}} \,
e^{ (5 - 2(S + J) ) A ( z )}
V(z)
\psi ( q,z ) \phi_a \phi_b .
\end{equation}
We denoted the non-normalizable mode for the current operator by
$\psi$, the normalizable modes by $\phi_{a,b}$ and the suitable
five-dimensional coupling constant by $g_d$.

The form factor must satisfy a constraint $F_{ab}(q^2=0)=\delta_{a
  b}$.  In our context, this is related to the
proper normalization of the
non-normalizable mode $\psi$. When the gauge theory is
conformal, so that the five-dimensional spacetime is
$AdS_5$, the usual choice of
normalization in the \adscft\ context is 
\begin{equation}\label{eq:normconformal}
\lim_{q^2 \to 0} (z/R)^{2(S-1)}
\psi(q^2,z) = \lim_{z\to 0} (z/R)^{2(S-1)} \psi(q^2,z) = 1.
\end{equation}
For this reason, we present the non-normalizable modes in this paper 
with the similar normalization,
\begin{equation}\label{eq:normnonconf}
\lim_{q^2 \to 0} (z/R)^{2(S-1)} e^{-2 (S-1) A(z)} \psi(q^2,z) = 1,
\end{equation}
which reduces to Eq.~(\ref{eq:normconformal}) in the ``conformal
limit'' $z \to 0$.  However, whenever we use these modes in the
computation of form factors, we need to make them ``canonically
normalized.'' This is accomplished by scaling $\psi \to \psi/g_d$,
which ensures that Eq.~(\ref{eq:formfactor}) reduces to
Eq.~(\ref{eq:normal}) in the $q^2\to 0$ limit, and that
$F_{ab}(q^2=0)=1$.

We can compute another quantity, which corresponds to a hadron
coupling constant among three hadron states. It is the three hadron
overlap, obtained in the following way.  When the hadron states
are labeled by $n$, $a$ and $b$, the three hadron coupling is given by
\begin{equation}
  g_{n a b} = g_d R^{5-2(S+J)}\int \frac{dz}{z^{5-2(S+J)}} \,
e^{ (5 - 2(S + J)) A ( z )}
V(z)
\varphi_n \phi_a
\phi_b.\label{eq:THO} 
\end{equation}

\section{Review of the hard-wall and the D3/D7 model}\label{sec:review-models}
Here we add a brief explanation of the models that we used for
examples and list the mode functions. 

As explained earlier, the hard-wall model is given by the
$\ads{5}\times S^5$ space with a wall at a finite radius. This wall
puts the boundary condition the mode functions and we choose the
Neumann condition. The metric is just given by the $\ads{5}\times S^5$
metric, $A(z)=0$ in Eq~(\ref{eq:metric}), up to the location of the
wall, $z=z_{max}=1/\Lambda$.
Then the spin one normalizable mode
corresponding to the current operator is 
given by 
\[
A_{\mu}(m_n)=\epsilon_\mu \phi_n(z){\cal Y}^0(S^5),\qquad {\cal Y}^0(S^5)=1, 
\]
\be\label{Ahadwave}
\phi_n(z) = \frac{\sqrt{2} z/z_{max} J_1 (
  \zeta_{0;n} z/z_{max})}{\pi^{\frac32} R^3 J_1 ( \zeta_{0;n} )} 
\ ,\ee
where $\zeta_{k;n}$ denotes the $n$-th zero of the Bessel function
$J_k(x)$. We also expressed the mode in terms of the $v$ coordinate
that we introduce below for the D3/D7 system. The mass is
\be\label{Ahadmass} m_n =
{\zeta_{0;n}}{\Lambda}\underset{\quad n\gg
1}{\longrightarrow} \left(n-\frac14\right)\pi {\Lambda} \
. \ee
The corresponding non-normalizable mode is
\[
\hat A_{\mu}(m_n)=\epsilon_\mu \psi(q,z),
\]
\begin{equation}\label{eq:Ahadnorm}
\psi(q,z) = 
q z\left\{
  K_1(qz)+\frac{K_0(q/\Lambda)}{I_0(q/\Lambda)} I_1(qz)\right\}
\underset{\Lambda\to0}{\approx} 
q z K_1(q z).
\end{equation}
Canonical normalization for this mode requires dividing
by $g_d=g_{10}=\kappa/R$, where
$\kappa^2=(2\pi)^{7}{\alpha'}^4 g_s^2/2$ and $R^4=4\pi g_s N
{\alpha'}^2$; thus $g_{10}=2 \pi^{5/2} R^3/N$.  The volume of
the internal manifold is that of a 5-sphere
of constant radius $R$: $V(z)=\pi^3 R^5$.

The spin-two case is similar. The normalizable and non-normalizable
modes are
\begin{eqnarray}
h_{\mu\nu} &=& \epsilon_{\mu\nu}
\frac{\sqrt{2} J_2(\zeta_{1,n} z/z_{max})}{\pi^\frac32
z_{max} R^2 J_2(\zeta_{1;n})}{\cal Y}^0(S^5),
\label{eq:hhadwave}\\
\hat h_{\mu\nu} &=& \epsilon_{\mu\nu} \frac{R^2q^2}{2}\left\{ K_2(q z) +
  \frac{K_1(q/\Lambda)}{I_1(q/\Lambda)} I_2(q z)\right\}{\cal Y}^0(S^5).\label{eq:hhadnorm}
\end{eqnarray}

For a scalar hadron created by an operator with conformal dimension
$\Delta$, we use the following mode
\begin{equation}\label{eq:shadnorm}
\phi^{(\Delta)}_n(z)=\frac{\sqrt{2} z^2
  J_{\Delta-2}(\zeta_{\Delta-3;n}z/z_{max})}{\pi^{\frac32}R^4 z_{max}
  J_{\Delta-2}(\zeta_{\Delta-3;n})},
\end{equation}
which satisfies the boundary condition
\[
\left. \partial_{z} \left[ z^{\Delta-4} \phi^{(\Delta)}_n(z)\right] 
\right|_{z=z_{max}} = 0,
\]
which is analogous to Neumann condition for the vector and rank two tensor 
modes.

The D3/D7 model is described in detail in~\cite{KK,KKW,Myers,hys}. Here we
summarize only what is needed in the computations.The theory is
composed of two sectors, \nfour\ $SU(N)$ Yang-Mills theory and $N_f$
of \ntwo\ hypermultiplets which are in fundamental representation of
$SU(N)$. It has the global symmetry $SO(4)\approx SU(2)\times SU(2)$
symmetry, consisting of an $SU(2)_\Phi$ symmetry rotating $\Phi_1$ and
$\Phi_2$ and an $SU(2)_R$ \ntwo\ R-symmetry.
The superpotential is
$$
W= \sqrt{2}\ {\rm tr}\Big([\Phi_1,\Phi_2]\Phi_3\Big) +
\sum_{r=1}^{N_f} Q^r\Phi_3\tilde Q_r + m_r Q^r\tilde Q_r
$$
where $m_r$ is the mass of hypermultiplet $r$ and the trace is
over color indices. If all the masses $m_r$ are equal, as we will
assume throughout, there is additional flavor symmetry $SU(N_f)$.

For large $g^2 N$, and in the ``quenched limit'' $N_f \ll N$, 
the theory is
dual to  IIB supergravity in $\ads{5}\times S^5$ with $N_f$ probe
D7 branes~\cite{KK}. The induced metric on the D7 brane is given
by
\bea d s^2 &=&  {r^2\over R^2} \eta_{\mn}dx^\mu dx^\nu +
\sum_{c=4}^7 {R^2\over r^2} (dx^c)^2 \cr &=&
\frac{L^2}{R^2}(\varrho^2 + 1) \eta_\mn dx^\mu dx^\nu + R^2
\frac{1}{\varrho^2 + 1} d \varrho^2 +R^2
\frac{\varrho^2}{\varrho^2 + 1} d \Omega_3^2 \ , \eea where
$\varrho^2= \frac{r^2}{L^2}-1$, and the $S^3$ involves the angular
coordinates in the four-dimensional space spanned by $x^4,
x^5,x^6,x^7$.

All of the Born-Infeld modes on the D7 brane were exactly calculated in
\cite{Myers}, where the modes were classified as scalar, I$\pm$, II
and III.  There are extra degeneracies, not explained by the explicit
global symmetries, among the scalar, II and III modes. This signifies
the existence of an extended $SO(5)$ accidental symmetry, containing
the original $SO(4)$ inside.

In~\cite{hys}, a set of coordinates was introduced such that
the expressions for the mode functions become more convenient for form
factor computation than those presented in~\cite{Myers}. 
These the
coordinates are
\bel{coords} v = (L/r)^2 \ ,  \ w = 1-v\  ;\ \varrho^2 = v^{-1}-1
= \frac{w}{1-w} \ . \ee

Among the various quarkonium modes in the theory, we only present the
mode I- and II, which we mainly used in this paper. The detailed
discussion on other modes can be found in~\cite{Myers}.

{\bf Type I-:} The I- modes correspond to the D7 brane worldvolume
Yang-Mills field polarized in the $S^3$ directions. Using the global
charges and the conformal dimension, they can be uniquely identified as 
dual to the operators
\bel{QphiQ} ( \tilde{Q} \Phi^{\elll-1} Q )_{\theta, \bar{\theta} =
0} = \tilde{Q} \Phi^{\elll-1} Q +
   \cdots
\ee
The masses of the normalizable mode are
$$
M_{I-}^2 = 4 m_h^2 (n+\elll
)(n+\elll +1) \ .
$$
The wavefunctions are
\bel{TypeI} A_{\mu} = 0,\quad A_{\rho} =
0,\quad A_{\alpha}\ = \phi^{I-} ( \rho ) e^{\ i k \cdot x}
   \mathcal{Y}_{\alpha}^{\elll ,-} ( S^3 )
\ee
\begin{eqnarray*}
\phi^{I-}_{\elll ,n}  & = & (C^{I}_{\elll n}/L) \varrho^{\elll +1}
(1+\varrho^2
)^{-1-n-\elll } F(-n, 1-n-\elll  ;\elll +2 ; -\varrho^2)\\
& = & (\hat C^{I}_{\elll n}/L) v^{(\elll  + 1) / 2} ( 1 - v )^{(
\elll  + 1 ) / 2}
   P_n^{( \elll  + 1, \elll  - 1 )} ( 2 v - 1 )
\end{eqnarray*}
where $n\geq 0$, $\elll \geq 1$, and
$$
C_{\elll n}^{I}= 
\frac1\pi\sqrt{(2n+2\elll+1) {{n+2\elll}\choose{\elll+1}}
{{n+\elll+1}\choose{\elll+1}} } =
 {{n+\elll+1}\choose{\ell+1}}
\hat C_{\ell n}^I \ .
$$

{\bf Type II:} These modes
correspond to the worldvolume gauge field polarized in 0123
directions, and are dual to the flavor current operator and
its generalizations,
\bel{eq:QphidQ}
 ( Q^{\dag} \Phi^\elll  Q - \tilde{Q} \Phi^\elll  \tilde{Q}^{\dag} )_{\theta
   \bar{\theta}} = Q^{\dag} \Phi^\elll  \partial^{\mu} Q + \psi^{\dag}_Q \Phi^\elll
   \sigma^{\mu} \psi_Q - \tilde{Q} \Phi^\elll  \partial^{\mu} \tilde{Q}^{\dag} +
   \cdots
\ee
where $\Phi^\elll $ stands for any product of $\Phi_1$ and
$\Phi_2$ which is a symmetric and traceless representation under
$SO(4)$.
The masses of the normalizable mode are
$$M_{II}^2 = 4 m_h^2 (n+\elll +1)(n+\elll +2) \ ,
$$
The wavefunctions are
\bel{TypeII} A_{\rho} = 0 \ ,\quad A_{\alpha} = 0 \ ,\quad A_{\mu}
= \zeta_{\mu} \phi^{II} ( \rho ) e^{i k \cdot x}
   \mathcal{Y}^{\elll } ( S^3 ) \ ,\quad k\cdot \zeta =0
\ee
\begin{eqnarray*}
\phi^{II}_{\elll ,n} & = & (C_{\elll n}^{II}/R^2) \varrho^{\elll
}(1 + \varrho^2)^{-1-n-\elll } F
(-n, -1-n-\elll ; \elll  + 2 ; - \varrho^2 )\\
& = & (\hat C_{\elll n}^{II}/R^2) v^{(\elll+2) /2} ( 1 - v
)^{\elll  / 2} P_n^{(\elll  + 1,\elll+1)} ( 2 v - 1 )\ ,
\end{eqnarray*}
where $n\geq0$, $\elll \geq 0$,
$$C_{\elll n}^{II}= 
\frac1\pi\sqrt{(2n+2\elll +3)
{{n+2\elll +2}\choose{\elll +1}} {{n+\elll +1}\choose{\elll +1}} }
= {{n+\elll+1}\choose{\elll+1}} \hat C_{\ell n}^{II} \ ,
$$
and $P_n^{( \alpha , \beta )} ( x )$ denotes a Jacobi polynomial. In
addition to the normalizable mode, the non-normalizable mode of type
II is dual to the flavor current operator. It is
\begin{equation}
\psi^{II}(q)=\frac{\pi\alpha (\alpha + 1)}{\sin\pi\alpha}
{}_2F_1(-\alpha,\alpha+1,2,1-v)
\end{equation}
where $\alpha = (-1+\sqrt{1-(q/m_h)^2})/2$.
Again, canonical normalization
requires dividing by $g_d=g_8$, the Yang-Mills
coupling in eight dimensions: 
$g_8=(2 \pi)^{5/2} g_s^{1/2}\alpha' = 2\sqrt{2} \pi^2 R^2/\sqrt{N}$.
The volume of the 3-sphere is 
$V(w)=2\pi^2 R^3 w^{3/2}=2\pi^2 R^3 (1-v)^{3/2}$.


\section{Proof of decomposition formula}\label{sec:proof}

Assume that the spacetime is asymptotically $\ads{5}\times W$, with
metric is given as Eq.~(\ref{eq:metric}). We consider a
five-dimensional gauge field $C_\mu = \epsilon_\mu e^{iq\cdot x}
\chi(q^2,z)$, where $z$ is the five dimensional radial coordinate
defined in appendix~\ref{sec:methodology}.  The normalizable mode is
$\phi_n(z)\propto\chi(-m_n^2,z)$ at $q^2=-m_n^2$, and the
non-normalizable mode is $\psi(q^2,z)\equiv\chi(q^2,z)$ for arbitrary
$q^2$.  With gauge $C_z = 0,\ q\cdot C = 0$, the action for
$\psi(q^2,z)$ is
\[
S = \int_0^{z_{max}}dz\, \left(\frac{R\ e^{A(z)}}{z}\right) 
V(z)
\left[ (\partial_z \psi)^2 + q^2 \psi^2 \right],
\]
which gives us the equation of motion
\[
\mathcal{L}\psi - q^2 \psi = 0, \quad
\mathcal{L} = \frac{e^{-A(z)} z}{R \ 
V(z)
}\partial_z 
\left(\frac{R\ e^{A(z)}}{z}
V(z)
\ \partial_z\right).
\]
Now the problem effectively reduces to that of a
field in a one-dimensional cavity. 
The linearity of the equation allows us easily to obtain 
Green's theorem:
\begin{multline}
\int_0^{z_{max}}dz\, \left(\frac{R\ e^{A(z)}}{z}\right) 
V(z)
\left [
\psi (\mathcal{L} - q^2 )\chi - \chi (\mathcal{L} - q^2 )\psi\right]\\
= -\lim_{z' \to 0} \left[ \psi(z') \mathcal{D}_{z'} \chi(z') -
\chi(z')\mathcal{D}_{z'}\psi(z') \right ], 
\end{multline}
where
\[
\mathcal{D}_{z'} = \frac{R\ e^{A(z')}}{z'}
V(z')
\ \partial_{z'}.
\]
We assumed here that there is no additional source at $z=z_{max}$, which is
automatically guaranteed by a Neumann boundary condition for the
gauge field at $z=z_{max}$.  This
implies that our solution $\psi(q^2,z)$, with Neumann boundary conditions 
at $z=0$, can be obtained as
\begin{equation}\label{eq:solvepsi}
\psi(q^2,z) = \psi(q^2,0) \lim_{z' \to 0} \frac{R}{z'}
V(z')
\partial_{z'}G(z,z';q^2)
\end{equation}
where we used $e^{A(z)} \to 1$ as $z \to 0$ in the
asymptotically $\ads{}$ space. $G(z,z';q^2)$ is the Green's function
satisfying the equation 
\begin{equation}\label{eq:greeneq}
\left(\frac{R\ e^{A(z)}}{z}\right) V(z)
(\mathcal{L} - q^2)G(z,z';q^2) = - 
\delta(z - z'),
\end{equation}
with Dirichlet boundary conditions.
Since the normalizable modes form a complete basis, we can construct
the Green's function as
\begin{equation}\label{eq:green}
G(z,z';q^2)=\sum_n \frac{\phi_n(z) \phi_n(z')}{q^2+m_n^2}.
\end{equation}
It can be easily checked that Eq.~(\ref{eq:green}) satisfies
(\ref{eq:greeneq}) by using the completeness relation,
\[
\left(\frac{R\ e^{A(z)}}{z}\right) V(z)
\sum_n \phi_n(z)\phi_n(z') = 
\delta(z - z').
\]
Hence, with ``canonical normalization''\footnote{See
  appendix~\ref{sec:methodology} for explanation of this convention.}
$\psi(q^2,z=0) = 1/g_d$, we obtain Eq.~(\ref{eq:drel}) with
(\ref{eq:fn}) by plugging (\ref{eq:green}) in (\ref{eq:solvepsi}). The
generalization to the other spin cases is straightforward and again
yields Eq.~(\ref{eq:drel}). Therefore, the decomposition~(\ref{Ffg})
is exact for every conserved current in the large $\lambda$ limit.

\end{document}